\documentstyle[12pt,epsfig,amssymb]{article}
\begin{document}
\setlength{\parskip}{0.45cm}
\setlength{\baselineskip}{0.75cm}
\begin{titlepage}
\begin{flushright}
CERN-TH/98-133 \\
ETH-TH/98-12 \\ 
DFPD  98/TH 19 \\
\end{flushright}
\vspace{0.1cm}
\begin{center}
\Large
{\bf The light--cone gauge and the calculation}

\vspace{0.1cm}
{\bf of the two--loop splitting functions}

\vspace{1.2cm}
\large
A.\ Bassetto$^a$, G.\ Heinrich$^b$, Z.\ Kunszt$^b$, W.\ Vogelsang$^c$ \\

\vspace*{1.5cm}
\normalsize
{\it $^a$Dipartimento di Fisica ``G.\ Galilei'', via Marzolo 8,
I--35131 Padova, Italy, \\ INFN, Sezione di Padova}

\vspace*{0.1cm}
{\it $^b$Institute of Theoretical Physics, ETH Z\"{u}rich, Switzerland}

\vspace*{0.1cm}
{\it $^c$Theoretical Physics Division, CERN, CH-1211 Geneva 23, Switzerland} \\
\vspace{1.5cm}
%
\large
{\bf Abstract} \\
\end{center}
\normalsize
We present calculations of next--to--leading order QCD splitting 
functions, employing the light--cone gauge method of Curci, Furmanski, and
Petronzio (CFP). In contrast to the `principal--value' prescription used in
the original CFP paper for dealing with the poles of the light--cone gauge 
gluon propagator, we adopt the Mandelstam--Leibbrandt prescription
which is known to have a solid field--theoretical foundation.
We find that indeed the calculation using this prescription 
is conceptionally clear and avoids the somewhat dubious manipulations
of the spurious poles required when the principal--value 
method is applied. We reproduce the well--known results for the 
flavour non--singlet splitting function and the 
$N_C^2$ part of the gluon--to--gluon singlet splitting function, which 
are the most complicated ones, and which provide an exhaustive test of the
ML prescription. We also discuss in some detail the $x=1$ endpoint 
contributions to the splitting functions.
\end{titlepage}
\newpage
\section{Introduction}
The advantages of working in axial gauges when performing perturbative QCD
calculations are known since a long time~\cite{do80}. 
Those gauges enable us to retain, in higher order calculations, 
a natural `partonic' interpretation for the vector field, typical
to leading log approximation.

Among axial gauges, the one which enjoys a privileged status is 
the light--cone axial gauge (LCA), 
characterized by the condition $n^\mu A_\mu=0$, 
$n^\mu$ being a light--like vector $(n^2=0)$. At variance with temporal 
$(n^2>0)$ and spacelike $(n^2<0)$ axial gauges, which do have problems 
already at the free level~\cite{basbook}, and with the spacelike 
`planar' gauge~\cite{do80} in which the behaviour of the theory in higher 
loop orders is still unsettled~\cite{basbook}, LCA  can be 
canonically quantized~\cite{bas1} and renormalized~\cite{bas} at all
orders in the loop expansion following a well--established
procedure. To reach this goal it is crucial to treat the `spurious' 
singularity occurring in the tensorial part 
of the vector propagator
\begin{equation}
\label{propml}
{\cal D}^{\mu\nu} (l) = \frac{i}{l^2+i \varepsilon} \Bigg( -g^{\mu \nu}+
\frac{n^\mu l^\nu+n^\nu l^\mu}{n l}  \Bigg) 
\end{equation}
according to a prescription independently suggested by 
Mandelstam~\cite{man} and Leibbrandt~\cite{leib} (ML) and  
derived in Ref.~\cite{bas1} in the context of equal--time 
canonical quantization:
\begin{equation}
\label{ML}
{1\over (nl)}\to {1\over [nl]}\equiv{1\over nl+i\varepsilon 
{\rm sign}(n^{*}l)}={n^{*}l\over 
n^{*}lnl+i\varepsilon} \; ,
\end{equation}
the two expressions being equal in the sense of the theory of
distributions. The vector $n^{*}$ is light-like and such that $n^{*}n=1$.
Denoting by $l_\bot$ the transverse part of the vector $l_\mu$, 
orthogonal to both $n_\mu$ and $n^{*}_\mu$, one has
\begin{equation}
2 (n l) (n^*  l) = l^2 + l_{\perp}^2 \; .
\end{equation}

The key feature of the ML prescription is that the spurious poles in the
complex $l_0$ plane are placed in the second and fourth quadrants,
i.e., with the same pattern as one encounters for usual covariant 
denominators like $1/(l^2+i \varepsilon)$. One can therefore perform a 
proper Wick rotation to Euclidean momenta, and a suitable power counting
criterion in the Euclidean integrals will give information on the 
ultraviolet (UV) 
divergencies of the corresponding Minkowskian integrals. This is in contrast 
to the Cauchy principal value (PV) prescription, which, under a Wick
rotation, entails further contributions and therefrom a violation of
power counting.

A crucial property of the ML distribution is the occurrence of two  
contributions with opposite signs in the absorptive part of the vector 
propagator~\cite{bastalk},
\begin{eqnarray}
\label{disc}
{\rm disc}[{\cal D}_{\mu\nu}(l)]&=&2\pi \, \delta(l^2)\Theta(l_0)
\left( -g_{\mu\nu}+{2n^{*}l\over n^{*}n}\cdot{n_\mu l_\nu+n_\nu l_\mu\over 
l^2_\bot} \right) \\ \nonumber
&&-2\pi \, \delta(l^2+l^2_\bot)\Theta(l_0){2n^{*}l\over n^{*}n}{n_\mu 
l_\nu+n_\nu l_\mu\over l^2_\bot} \; .
\end{eqnarray}
\noindent
Here the second, ghost--like, contribution (which is not present in the PV
prescription) is responsible for the milder infrared (IR) behaviour of the ML 
propagator. The presence of this axial ghost was stressed in~\cite{bas85};  
its properties are exhaustively discussed in~\cite{basbook}. Clearly, if one 
has a cut diagram with, say, $m$ final--state gluons, there is a 
discontinuity like (\ref{disc}) for each of the gluons, i.e., the phase space 
will split up into $2^m$ pieces. 

\def\Proj{{\cal P}}
\newcommand{\slsh}{\rlap{$\;\!\!\not$}}     
One of the most interesting and non--trivial applications of the LCA
is the computation of the (spin independent)
splitting functions for the two--loop Altarelli--Parisi (AP) evolution 
of parton densities, following a method proposed and used by Curci, 
Furmanski and Petronzio (CFP)
in Refs.~\cite{cfp,fp}. This method is based on the observation~\cite{egmpr} 
that in axial gauges the two--particle irreducible kernels of the ladder 
diagrams are finite, so that the collinear singularities that give rise to
parton evolution, only originate from the lines connecting the kernels. 
Therefrom, using renormalization group techniques, the splitting functions 
are obtained by some suitable projection of the ladder diagrams, exploiting 
the factorization theorem of mass singularities~\cite{egmpr}. 
We refrain from giving further details of the CFP method, since these can be
found in~\cite{cfp,ev}. We just mention at this point that one projects on  
the quantity $\Gamma_{ij}$, given by
\begin{equation} \label{gam1}
\Gamma_{ij} \left( x,\alpha_s,\frac{1}{\epsilon} \right)=
Z_j \Bigg[\delta(1-x) \delta_{ij} +x \; 
\mbox{PP} \int \frac{d^dk}{(2 \pi)^d}
\delta \left( x-\frac{nk}{pn} \right) 
U_i K \frac{1}{1-\Proj K} L_j \Bigg] \; , 
\end{equation}
where $2\epsilon=(4-d)$ and $K$ is a 2PI kernel, which is finite in the 
light--cone gauge~\cite{egmpr,cfp}. The labels $i,j$ run over quarks and
gluons; in the flavour non--singlet case one has just $\{ ij \} = \{ qq \}$.
Furthermore, in~(\ref{gam1}) PP denotes the pole part, and the 
projectors $U_i$,$L_i$ are given by
\begin{eqnarray}
&& U_q =\frac{1}{4 n k}\slsh{n}\; , \;\;\; L_q = \slsh{p} \; , \nonumber \\
&& U_g = -g^{\mu\nu} \; ,\;\;\; 
L_g = \frac{1}{d-2} \Bigg[-g^{\mu \nu}+\frac{n^\mu p^\nu+n^\nu p^\mu}{pn}
\Bigg] \; . 
\end{eqnarray}
The splitting functions $P_{ij}$ to the desired order can be read off
from the $1/\epsilon$ pole of $\Gamma_{ij}$:
\begin{equation} \label{gam}
\Gamma_{ij} \left( x,\alpha_s,\frac{1}{\epsilon} \right) = \delta (1-x) 
\delta_{ij} - 
\frac{1}{\epsilon} \Bigg(\frac{\alpha_s}{2\pi} P_{ij}^{(0)}(x)+
\frac{1}{2} \left( \frac{\alpha_s}{2\pi}\right)^2 P_{ij}^{(1)} (x) + 
\ldots \Bigg) + O \left(\frac{1}{\epsilon^2} \right) \; .
\end{equation} 
For future reference, we write down a similar expression~\cite{cfp} for 
the residue $Z_j$ ($j=q,g$) of the pole of the full quark (or gluon) 
propagator:
\begin{equation} \label{zq}
Z_j = 1 - \frac{1}{\epsilon} \Bigg(\frac{\alpha_s}{2\pi} \xi_j^{(0)}(x)+
\frac{1}{2} \left( \frac{\alpha_s}{2\pi}\right)^2 \xi_j^{(1)} (x) + 
\ldots \Bigg) + O \left(\frac{1}{\epsilon^2} \right) \; .
\end{equation} 
Inspecting Eqs.~(\ref{gam1}),(\ref{gam}),(\ref{zq}), we see that $Z_q$ and 
$Z_g$ contribute to the endpoint ($\sim \delta (1-x)$) parts of the 
splitting functions $P_{qq}$ and $P_{gg}$, respectively.

In the above references~\cite{cfp,fp}, the `spurious' singularity $1/nl$ 
of the gluon propagator was handled according to the PV prescription. 
The method of~\cite{cfp,fp} has been very successful in providing the 
first correct result for the next--to--leading order (NLO) gluon--to--gluon 
splitting function. The result previously obtained in the operator product 
expansion (OPE) method~\cite{ope2} was not correct due to a subtle conceptual
problem which was recently clarified~\cite{hvn,cs}. The new Feynman gauge 
OPE calculations confirmed the old CFP result. Despite of this success, 
the LCA calculation with PV prescription is considered dubious because of the
difficulties with power counting and Wick rotation mentioned above.
In particular it is not clear whether its `calculational rules'  remain 
valid in higher orders. We note that the precise description of some of the 
new high precision collider data call for the extension of the
NLO QCD analysis to {\em next}--to--next--to--leading order (NNLO).
Therefore a deeper understanding of the formal field--theoretical
basis of the CFP method is strongly motivated. The use of PV is by no means 
mandatory in the CFP method, it can also be applied when handling the 
$1/ln$ singularities with the theoretically more sound ML prescription. 

A first attempt using 
the ML prescription in connection with the CFP method
has been performed in Ref.~\cite{bastalk}, where the one-loop
AP splitting functions~\cite{ap} have been correctly
reproduced, both for the flavour non--singlet and for the flavour 
singlet case. A new characteristic feature of this calculation
is that `real' and `virtual' contributions are {\it separately}
well--defined in the limit $x\to 1$, $x$ being the longitudinal 
momentum fraction, at variance with the corresponding PV result.
This occurs thanks to the presence of the `axial' ghost,
which, standing by the usual gluon term, protects its singular 
behaviour with respect to the transverse momentum.
There is no need of any IR cutoff to regularize intermediate results.

Beyond one loop, the calculation of the splitting functions according to 
the CFP method in LCA with the ML prescription, has already been 
tackled in a recent paper~\cite{gz}. We believe, however, that improvements 
to the calculation~\cite{gz} can and should be made. First of all, only
the $C_F^2$ part of the flavour {\em non}--singlet splitting function is 
studied in~\cite{gz}. In this paper we will also calculate the $C_F T_f$ 
part and, in particular, the far more complicated piece $\sim C_F N_C$ of this 
function, as well as the $N_C^2$ part of the gluon--to--gluon splitting
function contributing to the flavour {\em singlet} sector. As we will show,
this set of functions we consider comprises all possible one--loop 
structures of QCD and thus enables an exhaustive test of the ML prescription
in this application. The ML 
calculation of the other singlet splitting functions,
like the non--diagonal quark--to--gluon (and vice versa) one, is therefore
not really required in this context: they will certainly come out correctly if
the prescription works for the far more complicated cases we study. 

Secondly, the power and virtues of the ML prescription were not 
fully exploited in~\cite{gz}, where some contributions resulting 
from the axial ghosts of the ML prescription were neglected. 
These contributions are $\sim \delta (1-x)$ and thus only affect the endpoints
of the diagonal splitting functions. Nevertheless, their inclusion is 
required for a complete analysis, since only then the crucial question 
of the finiteness of the two--particle--irreducible (2PI) kernels in 
the light--cone gauge can be fully answered. We also remind the reader in 
this context that in the original CFP papers~\cite{cfp,fp}, the endpoint 
contributions to the diagonal splitting functions were never determined 
by explicit calculation, but were derived in an indirect way from fermion 
number and energy--momentum conservation. The fact that we pay more attention
to the point $x=1$ will enable us to improve this situation to a certain 
extent: for the first time within the CFP method, we will determine the 
full part $\sim C_F T_f  \delta (1-x)$ of $P_{qq}^{V,(1)}$ by explicit 
calculation. 

Finally, in~\cite{gz} a principal value regularization
was still used at some intermediate steps of the calculation. Even though
this was only done at places where it seemed a safe and well--defined
procedure, it is more in the spirit of the ML prescription to abandon
the PV completely and to stick to one single regularization, the 
dimensional one. This view is corroborated by the observation that the
PV regularization as used in~\cite{gz} actually turns out to become 
technically too complicated when one studies the $C_F N_C$ part of the 
flavour non--singlet splitting function, or the $N_C^2$ part of 
$P_{gg}^{(1)}$. 

The remainder of this paper is organized as follows: to set the
framework, we will present a brief rederivation of the leading order (LO)
quark--to--quark splitting function $P_{qq}^{(0)}$ in sec.~2. Section~3 
will contain the  
calculation of the flavour non--singlet splitting function at two loops.
More specifically, we will discuss in detail the treatment of the 
various virtual--cut and real--cut contributions in subsections~3.1 and 3.2,
respectively, while sec.~3.3 presents the final results of the calculation.
In~3.4, we discuss the endpoint contributions and provide a sample 
calculation of a two--loop contribution to the quark self--energy 
in the LCA with ML prescription. Section~4 deals with the calculation
of the $N_C^2$ part of $P_{gg}^{(1)}$. Finally, we summarize our 
work in sec.~5.  
\section{Recalculation of the LO splitting function}
As a first example, we will rederive the LO result for the flavour 
non--singlet splitting function, using the ML prescription. 
This is a rather trivial calculation that nevertheless displays the main
improvements provided by the use of ML. Furthermore, the virtual graphs
in the NLO calculation have the LO kinematics, so this section also 
serves to prepare the NLO calculation. We noted before that
the LO example has already been worked out in~\cite{bastalk} where
collinear poles were regularized by taking the initial quark off--shell,
$p^2<0$, rather than by using dimensional regularization. This is perfectly 
fine at the LO level, but beyond LO it seems a forbidding task to keep
$p^2 \neq 0$, and in fact the underlying method of~\cite{cfp,fp} that we 
are employing has been set up in such a way that it relies on the use of 
dimensional regularization, yielding final results that correspond to
the $\overline{\rm{MS}}$ scheme. It therefore seems a useful exercise
to sketch the calculation of $P_{qq}^{(0)}$ in the ML prescription 
if dimensional regularization is used. 

The Feynman diagrams contributing to $\Gamma_{qq}$ at LO are shown in
Fig.~1. For the gluon polarization tensor in diagram (a) we need to insert
the two parts of the ML discontinuity in~(\ref{disc}) with
their two different $\delta$--functions. The first part of the
phase space, resulting from $\delta (l^2)$, can be written as
\begin{equation} \label{ps1}
x \int d^d k \int d^d l \; \delta \left( x-\frac{kn}{pn} \right) 
\delta (p-k-l) \delta (l^2) = \frac{\pi^{1-\epsilon}}{2 \Gamma(1-\epsilon)}
\int_0^{Q^2} d|k^2| (k_{\perp}^2)^{-\epsilon} \; ,
\end{equation}
where
\begin{equation}
k_{\perp}^2 = l_{\perp}^2 = |k^2| (1-x) \; .
\end{equation}
The $\delta (l^2)$ contribution of graph (a) to $\Gamma_{qq}$ is then given by
\begin{equation} \label{ddel}
\Gamma_{qq}^{(a),\delta (l^2)} (x) = \frac{\alpha_s}{2\pi} 
{\rm PP} \int_0^{Q^2}d|k^2| |k^2|^{-1-\epsilon} 
\; C_F (1-x)^{-\epsilon} \frac{1+x^2}{1-x} \; .
\end{equation}
Using the identity
\begin{equation} \label{del}
(1-x)^{-1-\epsilon} \equiv -\frac{1}{\epsilon} \delta (1-x)
+ \frac{1}{(1-x)_+} - \epsilon \left( \frac{\ln (1-x)}{1-x} \right)_+
+{\cal O} (\epsilon^2) \; ,
\end{equation}
\begin{figure}[hb]
\vspace*{-1cm}
\hspace*{-0.4cm}
\epsfysize38cm
\leavevmode\epsffile{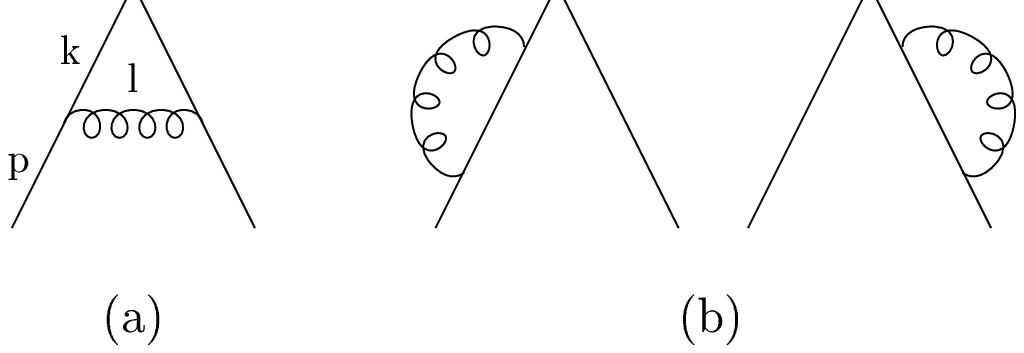}
\vspace*{-31.5cm}
\caption{\label{fig1}{\sf Diagrams contributing to $\Gamma_{qq}$ at LO.}}
\end{figure}
where the `plus'--prescription is defined in the usual way, one readily obtains
\begin{equation} \label{pa1}
\Gamma_ {qq}^{(a),\delta (l^2)} (x) = \frac{\alpha_s}{2\pi} 
{\rm PP} \int_0^{Q^2} d|k^2| |k^2|^{-1-\epsilon} 
\; C_F \left[ -\frac{2}{\epsilon}
\delta (1-x) + \frac{1+x^2}{(1-x)_+} \right] \; .
\end{equation}
For the ghost--like part we introduce the variable $\kappa$ as
\begin{equation} \label{def2}
k_{\perp}^2 = l_{\perp}^2 = -l^2 = |k^2| \kappa \; .
\end{equation}
The phase space is then given by
\begin{eqnarray} \label{ps2}
\int d^d k \hspace*{-0.1cm} \int d^d l \; 
\delta \left( x-\frac{kn}{pn} \right) \delta (p-k-l)
\delta (l^2+l_{\perp}^2) && \nonumber \\
&&\hspace*{-6cm} = \frac{\pi^{1-\epsilon}}
{2 \Gamma(1-\epsilon)} \int_0^{Q^2} d|k^2| \int_0^1 d\kappa 
\left( k_{\perp}^2 \right)^{-\epsilon} 
\delta \left((1-x)(1-\kappa) \right)
\nonumber \\
&&\hspace*{-6cm} = \frac{\pi^{1-\epsilon}}
{2 \Gamma(1-\epsilon)} \delta (1-x) \int_0^{Q^2} 
d|k^2| \int_0^1 \frac{d\kappa}{1-\kappa} 
\left( k_{\perp}^2 \right)^{-\epsilon} \; ,
\end{eqnarray}
where the last line follows since the root of the delta function
for $\kappa=1$ never contributes when we insert the second term 
in~(\ref{disc}) for the gluon polarization tensor into graph (a), thanks to 
the factor $2n^*l/l_{\perp}^2 \sim (1-\kappa)/\kappa$ accompanying the $\delta 
(l^2+l_{\perp}^2)$ in~(\ref{disc}). Thus, the ghost part contributes 
only at $x=1$. The contribution $\sim 1/\kappa$ of $2n^*l/l_{\perp}^2$
gives rise to a $1/\epsilon$--pole in the final answer:
\begin{equation} \label{pa2}
\Gamma_{qq}^{(a),\delta (l^2+l_{\perp}^2)} (x) = \frac{\alpha_s}{2\pi} 
\delta (1-x) {\rm PP} \int_0^{Q^2} d|k^2| |k^2|^{-1-\epsilon} \; C_F \left[
\frac{2}{\epsilon} \right] \; .
\end{equation}
Adding Eqs.~(\ref{pa1}) and (\ref{pa2}), we get the full contribution
of graph (a) to $\Gamma_{qq}$:
\begin{equation} \label{pa}
\Gamma_{qq}^{(a)} (x) = \frac{\alpha_s}{2\pi} 
{\rm PP} \int_0^{Q^2} d|k^2| |k^2|^{-1-\epsilon} 
\; C_F \frac{1+x^2}{(1-x)_+} \; .
\end{equation}
An important feature of this result should be emphasized, as it will also
be encountered at NLO: the integrand in (\ref{pa}) is completely finite, 
in a distributional sense. In other words, using the ML
prescription, we have verified the finiteness of the LO 2PI kernel 
$q\rightarrow qg$ in the light--cone gauge. We point out, however, that 
the finite 2PI kernel arises as the sum of the two {\em singular} pieces 
in Eqs.~(\ref{pa1}),(\ref{pa2}). This is again a finding that will 
recur at NLO: the full discontinuity~(\ref{disc}) of the gluon propagator 
in the ML prescription has a much milder behaviour than the 
individual contributions to it.

It is instructive to contrast the result in~(\ref{pa}) with the one 
obtained for the PV prescription~\cite{cfp}:
\begin{equation} \label{papv}
\Gamma_{qq}^{(a),PV} (x) = \frac{\alpha_s}{2\pi} 
{\rm PP} \int_0^{Q^2} d|k^2| |k^2|^{-1-\epsilon} 
\; C_F \Bigg[ \frac{1+x^2}{(1-x)_+} + 2 I_0 \delta (1-x) \Bigg] \; ,
\end{equation}
where 
\begin{equation} \label{io}
I_0 \equiv \int_0^1 \frac{u}{u^2+\delta^2} du \approx -\ln \delta \; .
\end{equation}
Thus, the 2PI kernel for the PV prescription has a divergent coefficient
of $\delta (1-x)$, resulting from the gauge denominator $1/nl$ and being
regularized by the parameter $\delta$.

The calculation of the LO splitting function is completed by determining
the endpoint contributions at $x=1$, corresponding to $Z_q$ in~(\ref{gam1})
and given by the graphs in Fig.~1(b). They can be straightforwardly 
obtained\footnote{Alternatively one can obtain the contributions from the 
requirement $\int_0^1 P_{qq}^{(0)}(x) dx =0$ \cite{ap,cfp}.} using 
the UV--singular structure of the one--loop quark self--energy, determined 
for the ML prescription in~\cite{lny}. One finds~\cite{bastalk}:
\begin{equation}
Z_q = 1 - \frac{\alpha_s}{2\pi}\frac{1}{\epsilon} C_F \frac{3}{2} \; ,
\; \; \; \; \; \mbox{{\rm that is}}, 
\; \; \; \; \; \xi_q^{(0)}=\frac{3}{2}C_F \; .
\end{equation}
Putting everything together, one eventually obtains 
\begin{equation} \label{paqq}
P_{qq}^{(0)} (x) = C_F \Bigg[ \frac{1+x^2}{(1-x)_+} +\frac{3}{2} \delta 
(1-x) \Bigg] \; , 
\end{equation}
in agreement with~\cite{ap}. We finally note that of course the same 
{\em final} answer is obtained within the PV prescription: the singular 
integral $I_0$ in~(\ref{papv}) is cancelled by the contribution from 
$Z_q$, since we have~\cite{cfp}
\begin{equation}
Z_q^{PV} = 1 - \frac{\alpha_s}{2\pi}\frac{1}{\epsilon} C_F \Bigg[ 
\frac{3}{2} -2 I_0 \Bigg] \; .
\end{equation}
Thus, to summarize, the advantage of the ML prescription at the LO level 
mainly amounts to producing truly finite results for the 2PI kernels, 
as required for the method of~\cite{egmpr,cfp,fp}. Furthermore, there is 
no need for introducing renormalization constants depending on additional 
singular quantities like $I_0$ that represent a mix--up in the treatment
of UV and IR singularities.
\newpage
\section{The calculation of the flavour non--singlet 
splitting function at NLO}
At NLO, there are two different non--singlet 
evolution kernels, $P^{-,(1)}$ and $P^{+,(1)}$, governing the 
evolutions of the quark density combinations $q-\bar{q}$ and 
$q+\bar{q} -(q'+\bar{q}')$, respectively. The two kernels are given in terms
of the (flavour--diagonal) quark--to--quark and quark--to--antiquark splitting
functions by (see, for instance, Ref.~\cite{ev})  
\begin{equation}
P^{\pm,(1)} \equiv P_{qq}^{V,(1)} \pm P_{q\bar{q}}^{V,(1)} \; ,
\end{equation}
where the last splitting function originates from a tree graph that does
not comprise any real--gluon emission and is therefore free of any problems
related to the use of the light--cone gauge. Thus, we do not need to 
recalculate $P_{q\bar{q}}^{V,(1)}$. The Feynman diagrams contributing 
to $P_{qq}^{V,(1)}$ are collected in Fig.~\ref{fig2}. 
We have labelled the graphs according to the notation of~\cite{cfp,ev}. 
We also show the graphs contributing to $Z_q$ at 
two loops. We will not calculate all of these, since this is not really 
required. Their role will be discussed in subsection~3.4. 
\subsection{Virtual--cut diagrams and renormalization}
\begin{figure}[t]
\hspace*{0.7cm}
\epsfysize23cm
\leavevmode\epsffile{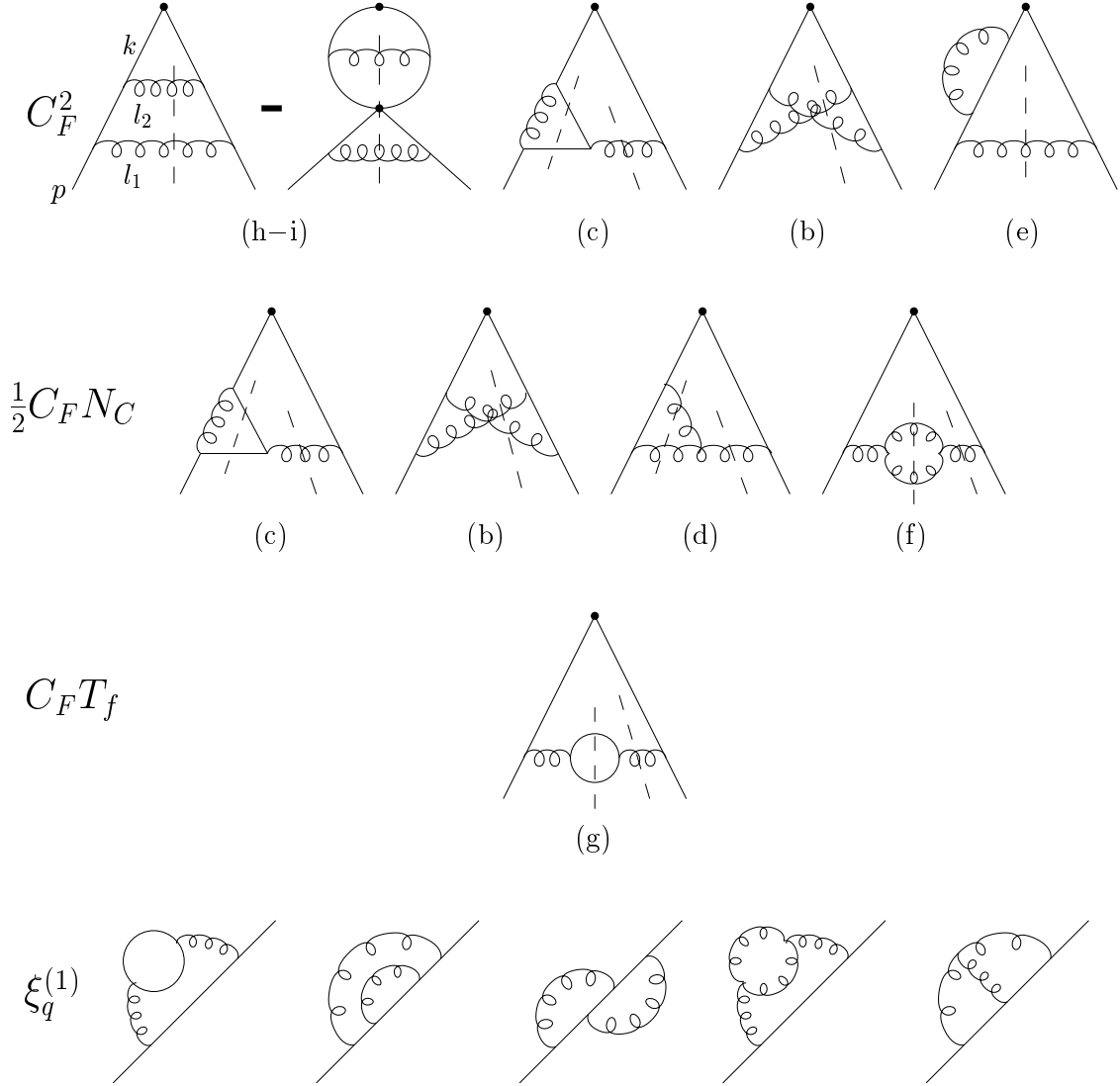}
\vspace*{-7cm}
\caption{\label{fig2}{\sf Diagrams contributing to $\Gamma_{qq}$ at NLO.}}
\end{figure}

Many of the diagrams in Fig.~\ref{fig2} have real and virtual cuts, 
as has been indicated by the dashed lines. Let us start by discussing
the contributions from the virtual cuts in graphs 
(c),(d),(e),(f),(g). It is clear that these essentially have the 
LO topology in the sense that 
there is always one outgoing gluon (momentum $l$), to be treated according 
to the ML prescription as discussed in the previous section. We recall 
that this means that there are two contributions for this gluon, one at 
$l^2=0$, and the other with $l^2+l_{\perp}^2=0$, corresponding to the 
gluon acting as an axial ghost\footnote{In the next subsection we will
see that for diagrams (d),(f) there are also other contributions 
at $l^2+l_{\perp}^2=0$, not just the one from the axial ghost 
going into the loop. However, the integration of those contributions 
proceeds in exactly the same way as outlined here. We postpone the
discussion of all the contributions at $l^2+l_{\perp}^2=0$ for graphs (d),(f)
to the next subsection.}. This immediately implies that we will have 
to calculate the loop integrals for these two situations. In addition, it is
clear that the ML prescription also has to be used in the calculation of 
the loop itself, not just for the treatment of the external gluon. For
instance, the gauge denominator $1/(r\cdot n)$, where $r$ is the loop
momentum, is subject to the prescription (\ref{propml}).
In short, we will need several two--point and three--point functions
with and without gauge denominators like $1/(n\cdot r)$, and for both
$l^2=0$ and $l^2+l_{\perp}^2=0$. 

We point out that important qualitative differences with respect 
to the PV prescription arise here: in the PV calculation one always 
has $l^2=0$ for the outgoing gluon in the virtual--cut graphs, and there 
is no explicit dependence on $l_{\perp}^2$. For instance, 
the way to deal with the self--energies in graphs (f),(g) in the PV
prescription is identical to their treatment in covariant gauges: one 
calculates them for off--shell $l^2$, renormalizes them, and eventually 
takes the limit $l^2\rightarrow 0$. In this way, almost all contributions 
of the diagrams will vanish since all loop integrals have to be 
proportional to $(l^2)^{-\epsilon}$ ($\epsilon <0$) on dimensional grounds. 
Only the contribution from the $\overline{\rm{MS}}$ counterterm 
remains~\cite{nowak,cfp} because this is the only quantity not proportional 
to $(l^2)^{-\epsilon}$. In contrast to this, in the ML prescription 
$\l_{\perp}^2$ sets an extra mass scale. For graph (f), one therefore 
encounters terms $\sim (l^2)^{-\epsilon}$, but also terms of the form 
$\sim (a \, l^2+b \, l_{\perp}^2)^{-\epsilon}$. The latter terms 
yield non--vanishing contributions to the virtual--cut result {\em even} 
at $l^2=0$. This is still not the case for graph (g) since here the pure 
quark loop of course does not contain any light--cone gauge denominator and 
thus does not depend on $l_{\perp}^2$. Nevertheless, one gets a 
contribution from the quark loop in (g) for $l^2+l_{\perp}^2=0$, i.e., 
when the gluon running into the loop is an axial ghost, corresponding 
to the second part of the ML discontinuity in~(\ref{disc}).

As expected, in the actual derivation of the loop integrals the 
property of the ML prescription to allow a Wick rotation is of great
help. Nevertheless, some of the integrals are quite involved, since the 
ML prescription introduces explicit dependence of the loop integrands
on the transverse components $r_{\perp}^2$ due to the identity
$2 (nr) (n^*r) =r^2+r_{\perp}^2$. Furthermore, since we are interested in 
calculating also the contributions at $x=1$, we need to calculate the 
loop integrals up to ${\cal O}(\epsilon)$ rather than ${\cal O}(1)$. 
The reason for this is that very often the final answer for a loop
calculation with $l^2=0$ will contain terms of the form
$(1-x)^{-1-a \epsilon}$ ($a=1,2$), 
to be expanded according to Eq.~(\ref{del}). As a 
result, a further pole factor $1/\epsilon$ is introduced into the calculation,
yielding finite contributions when multiplied by the ${\cal O}(\epsilon)$
terms in the loop integrals. A similar thing happens in the loop part 
with $l^2+l_{\perp}^2=0$. Here, an extra factor $1/\epsilon$ can be 
introduced when integrating this part over the phase space in~(\ref{ps2}). The 
higher pole terms created in these ways will cancel out eventually, 
but not the finite parts they have generated in intermediate steps of the
calculation. The detailed expressions for the loop integrals in the
ML prescription are given in Appendix~A. 

\begin{figure}[ht]
\hspace*{0.8cm}
\epsfysize24cm
\leavevmode\epsffile{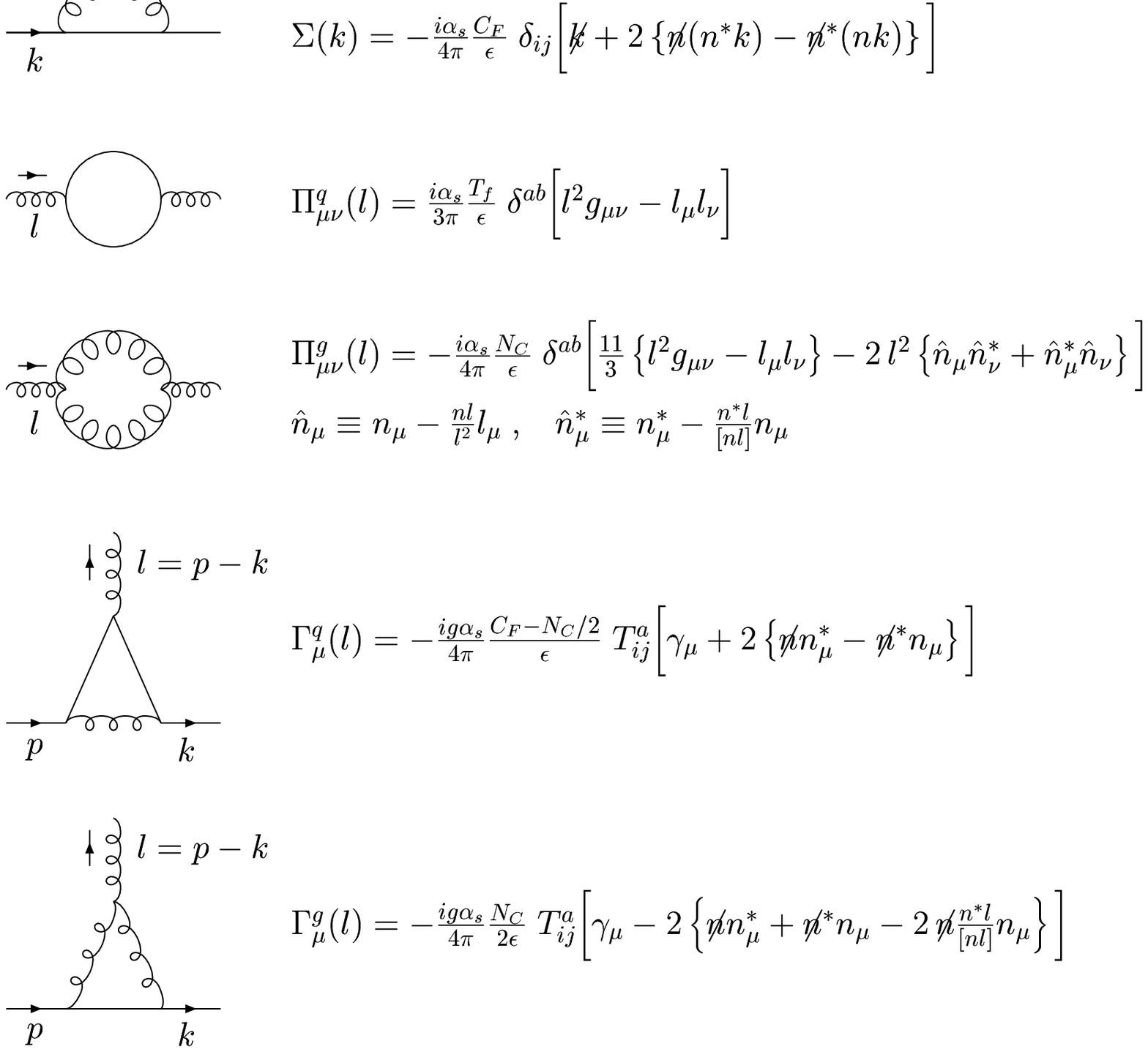}
\vspace*{-9.5cm}
\caption{\label{fig3}{\sf UV--divergent one--loop structures as obtained
in the light--cone gauge, using the ML prescription. The indices $i,j$ ($a,b$)
denote quark (gluon) colours; $T^a$ are the generators of SU(3).}}
\end{figure}

For the renormalization of the loop diagrams, one needs to subtract their 
UV poles, which is achieved in the easiest way 
by inserting the UV--divergent one--loop 
structures as calculated for the light--cone gauge in the ML prescription 
in~\cite{leib,lny,dal,lei}. All structures have also been compiled 
in~\cite{basbook}. The ones we need for the non--singlet calculation 
are displayed in Fig.~\ref{fig3}. One notices that, as
expected, the structures are gauge--dependent and Lorentz non--covariant.
Even more, the expressions for the non--Abelian quantities  $\Pi_{\mu\nu}^g$ 
and $\Gamma^g_{\mu}$ in Fig.~\ref{fig3} are non--polynomial in the external 
momenta, owing to terms $\sim 1/[nl]$. It is an important 
feature of the ML prescription that these non--local terms exist, 
but decouple from physical Green's functions~\cite{bas} thanks to the 
orthogonality of the free propagator with respect to the
gauge vector, $n_{\mu} {\cal D}^{\mu\nu} (l)=0$ (this has actually been an 
important ingredient for the proof~\cite{bas} of the renormalizability of QCD
in the ML light--cone gauge). Thus, the non--local pole parts never appear in 
our calculation. This is in contrast to the PV prescription, where one
has~\cite{cfp} contributions from the renormalization constants to the 
calculation that explicitly depend on the external momentum fractions $x$,
$1-x$.
\subsection{Real--cut graphs}
Let us now deal with the real cuts. One way of evaluating these 
is to integrate over the phase space of the two outgoing particles with
momenta $l_1$,$l_2$ (for the notation of the momenta, see Fig.~\ref{fig2}), 
in addition to the integration over the `observed' parton with momentum $k$.
This is the strategy we have adopted for all diagrams contributing 
to the $C_F^2$ and the $C_F T_f$ parts of the splitting function, 
i.e., graphs (b),(c),(g),(h). For graphs (d),(f), we found it simpler 
to use a different method, as will be pointed out below. 

If the two outgoing partons are gluons, their phase space in the ML
prescription splits up into four pieces, as we discussed in Sec.~1.
It is possible to write down a phase space that deals with all
four parts. We leave the details for Appendix~B.

Upon integration of the squared real--cut matrix element 
for a diagram, each of the four parts of phase space gives highly 
divergent results, but their sum is usually less singular. This is similar
to the pattern we found at LO. For instance, the phase space integration
of graph (b) of Fig.~2 (before performing the final integration over $|k^2|$) 
is expected to give a finite result, since the graph 
is 2PI and possesses no virtual cut. This indeed turns out to be the case,
but the individual contributions to (b) by the four parts of phase space 
all have poles $\sim 1/\epsilon^2$ and $\sim 1/\epsilon$ which cancel 
out when combined. A similar cancellation of higher pole terms happens 
for graph (h) (which of course is not finite by itself, but has a left--over 
$1/\epsilon$ singularity, to be cancelled by the contribution 
from diagram (i)). Here even poles $\sim 1/\epsilon^3$ occur at intermediate
stages of the calculation. For those graphs that also have virtual cuts,
the situation is in general even more complicated, as cancellations will occur
only in the sum of the real and virtual cuts. An example for this case will
be given in sec.~3.3. 

For graphs (d),(f), the phase space integrals become extremely 
complicated. This is due to the extra denominator $1/(l_1+l_2)^2$
present in these graphs, which causes great complications
in the axial--ghost parts of the phase space. For (f), we found it still 
possible to get the correct result via the `phase space method', but for
(d) this seemed a forbidding task. It turned out to be more convenient 
to determine the result in a different way: if one calculates, for instance,
the gluon loop in graph (f) for an arbitrary off--shell momentum $l$ going
into the loop, the imaginary part of the loop will correspond to the 
real--cut contribution we are looking for. To be more precise, the 
strategy goes as follows: we calculate the loop graph for off--shell 
$l$ and insert the result into the appropriate LO phase space. The latter 
can be derived as in (\ref{ps1}), omitting, however, the $\delta (l^2)$ 
there. One finds
\begin{equation} \label{ps11}
x \int d^d k \int d^d l \; \delta \left( x-\frac{kn}{pn} \right) \delta (p-k-l)
= \frac{\pi^{1-\epsilon}}{2 \Gamma(1-\epsilon)}
\int_0^{Q^2} d|k^2| \int_0^{1/(1-x)} d\tau \; (k_{\perp}^2)^{-\epsilon}\; ,
\end{equation}
where now
\begin{equation} \label{kdef}
k_{\perp}^2 = l_{\perp}^2 =|k^2| (1-x) \tau \; , \;\;\;
l^2 = \frac{|k^2| (1-x)}{x} (1-\tau) \; .
\end{equation}
The limits for the $\tau$ integration in~(\ref{ps11}) span the largest
possible range for $\tau$, given by the conditions $l_{\perp}^2>0$, 
$l^2+l_{\perp}^2>0$. The full imaginary part arising when performing the 
loop and the $\tau$ integrations has to correspond to the sum over all
cuts in the diagram. One encounters discontinuities from the following
sources:
\begin{description}
\item[(A)] 
from the loop integrations. Here imaginary parts arise, for instance,
if for certain values of $\tau$ and of the Feynman parameters 
$t_1,\ldots,t_k$, one finds terms of the form 
\begin{equation}
\Big( f(t_1,\ldots,t_k,\tau) \Big)^{-\epsilon} \; ,
\end{equation}
where $f$ is {\em negative}. Details for integrals with such properties are
given in Appendix~C. The imaginary part originating in this way essentially 
corresponds to the cut through the loop itself, i.e., to the real--cut 
contribution we are looking for.
\item[(B)] from the propagator $1/(l^2+i \varepsilon)$ via the identity
\begin{equation} \label{ip}
\frac{1}{l^2+i \varepsilon} = {\rm PV} \Bigg( \frac{1}{l^2} \Bigg) -
i \pi \delta (l^2) \; ,
\end{equation}
where PV denotes the principal value.
The imaginary part $\sim \delta (l^2)$ obviously represents the loop
contribution at $l^2=0$ which we have determined in the last 
subsection. Therefore, we do not need to reconsider this part of the 
discontinuity.
\item[(C)] from terms $\sim 1/[nl]$, for which a relation similar 
to~(\ref{ip}) holds, 
\begin{equation} \label{ip1}
\frac{1}{[nl]} \equiv \frac{1}{nl+i \varepsilon {\mbox {\rm sign}} (n^*  l)}
= {\rm PV} \Bigg( \frac{1}{nl} \Bigg) - i \pi {\mbox {\rm sign}} (n^*  l)
\delta (nl) \; .
\end{equation}
At first sight, one might think that the discontinuity $\sim \delta (nl)$ 
simply corresponds to the calculation of the gluon loop for the case when 
the gluon entering the loop is an axial ghost with $l^2=-l_{\perp}^2$.
However, the situation is more subtle: The terms $\sim 1/[nl]$ do not 
only originate from the propagators of the external gluons, but also 
from splitting formulas like~\cite{basbook} 
\begin{equation}
\frac{1}{[nr] [n(l-r)]} = \frac{1}{[nl]} \Bigg( 
\frac{1}{[nr]} + \frac{1}{[n(l-r)]} \Bigg) 
\end{equation}
(where $r$ is the loop momentum), as well as from the loop integrals 
themselves, like in the case of
\begin{equation}
\int \frac{d^d r}{(2\pi)^d} \frac{1}{(r^2+i \varepsilon) 
( (l-r)^2+i \varepsilon) [nr]} \; .
\end{equation}
All these terms $\sim 1/[nl]$ have to be treated according to the 
ML prescription, i.e., give rise to discontinuities $\sim \delta (nl) 
\sim \delta (1-x)$
via~(\ref{ip1}). The sum of all discontinuities arising in this way 
actually has to correspond to the `pure--axial--ghost' part of the graph, 
given by (a) the virtual--cut contribution when the gluon going into the 
loop is an axial ghost, {\em plus} (b) the real--cut contribution when 
{\em both} final--state gluons act as axial ghosts. These two parts cannot
easily be separated from each other, which is the reason why we  
postponed the whole treatment of graphs (d),(f) at $l^2=-l_{\perp}^2$ to 
this section. The integrals needed to obtain this part of the discontinuity
are those already mentioned in the last subsection and collected in the 
right--hand column of Tab.~4 of Appendix~A.  
\end{description}
It is also worth mentioning that despite the fact 
that graph (f) has a {\em squared} gluon propagator, there are 
cancellations coming from the algebra in the numerator; as a consequence, 
one never encounters expressions like $1/[nl]^2$ or 
$1/(l^2+i \varepsilon)^2$ before taking the discontinuity, 
and~(\ref{ip1}),(\ref{ip}) are all we need. 

Clearly, when finally collecting all the imaginary parts from (A) and (C), 
the PV--parts in (B) and (C) play a role in the calculation. While 
$1/nl \sim1/(1-x)$ in~(\ref{ip1}) only diverges at the endpoint at 
$x=1$ where it is always regularized by factors like $(1-x)^{-\epsilon}$,
the propagator $1/l^2$ in~(\ref{ip}) in general has its singularity inside
the region of the $\tau$ integration: from~(\ref{kdef}) one finds that
$l^2>0$ for $\tau<1$, but $l^2<0$ for $\tau>1$. The principal value 
prescription\footnote{To avoid confusion, we emphasize at this point that
the principal value for $1/l^2$ in~(\ref{ip}) is well--defined here and 
{\em not} related to the principal value prescription for the light--cone
denominator $1/nl$ that we heavily criticised in the introduction.} 
in~(\ref{ip}) takes care of the pole at $\tau=1$ and leads to a 
cancellation of the positive spike for $\tau \rightarrow 1^-$ and the 
negative one for $\tau \rightarrow 1^+$, resulting in a perfectly 
well--defined finite result.

The vertex graph (d) can be treated in a similar fashion as (f). Here
one calculates the full vertex for $p^2=0$, $k^2<0$, but arbitrary $l^2$,
and determines the imaginary parts arising with respect to $l^2$. 
This corresponds again to point (A) above, and calculational details 
are also given in Appendix~C. The imaginary part from (B) is again related 
to the virtual--cut contribution at $l^2=0$ that we already calculated in the 
last subsection. 
The discontinuity from (C) needs to be taken into account, and 
as before it corresponds to the full `pure--ghost' contribution (virtual--cut
{\em and} real--cut), residing at $x=1$.

A final comment concerns graph (i). Its contribution to $\Gamma_{qq}$ 
is essentially given by a convolution of two LO expressions, each 
corresponding to Fig.~1(a), keeping however also all {\em finite} terms
in the upper part of the diagram, including the factor $(1-x)^{-\epsilon}$ 
from phase space (see~(\ref{ddel}),(\ref{del})):
\begin{equation} \label{conv}
{\rm (i)} \sim \frac{1}{\epsilon} \Bigg[ \frac{1+z^2}{(1-z)_+}
-\epsilon (1+z^2) \Bigg( \frac{\ln (1-z)}{1-z} \Bigg)_+ - 
\epsilon (1-z) \Bigg] 
\otimes \frac{1}{\epsilon} \Bigg[ \frac{1+z^2}{(1-z)_+} \Bigg] \; ,
\end{equation}
where
\begin{equation} \label{conv1}
\Big( f\otimes g \Big) (x) \equiv \int_x^1 \frac{dz}{z} f\Big(\frac{x}{z}
\Big) g(z) \; .
\end{equation}
Note that this is in contrast to the PV prescription where the contribution
from (i) does not correspond to a genuine convolution in the mathematical 
sense. Since both of the convoluted functions in~(\ref{conv}) contain 
distributions, the convolution itself will also be a distribution. The 
evaluation of~(\ref{conv}) is most conveniently performed in Mellin--moment 
space where convolutions become simple products.  Some details of the 
calculation and a few non--standard moment expressions are given in Appendix~D.
\subsection{Final results}
We now combine the results of the previous subsections. The first observation
is that for the ML prescription all 2PI graphs, and also the difference 
(h$-$i), turn out to give truly finite contributions to $\Gamma_{qq}$, 
before the final integration over $|k^2|$ is performed. 
This expectation for the light--cone gauge~\cite{egmpr} was 
{\em not} really fulfilled by the PV prescription, where the
results for the diagrams depended on integrals like $I_0$ in (\ref{io})
that diverge if the regularization $\delta$ of the PV prescription is
sent to zero~\cite{cfp,ev}. 
The finiteness of the kernels in the ML prescription 
comes about via delicate cancellations of terms sometimes as singular as 
$1/\epsilon^2$ or even $1/\epsilon^3$ when the various real--cut 
(gluon and axial--ghost) and virtual--cut contributions are added. 
To give just one example beyond those already discussed in the previous 
subsection, let us discuss the contributions of graph (g) to $\Gamma_{qq}$. 
From the real--cut diagram, one has up to trivial factors
\begin{eqnarray} \label{pp1}
\Gamma_{qq}^{({\rm g}),r} &\sim& \left( \frac{\alpha_s}{2\pi}\right)^2 
{\rm PP} \int_0^{Q^2} \frac{d|k^2| |k^2|^{-1-2 \epsilon}}
{\Gamma (1-2 \epsilon )}
\Bigg[ \delta (1-x) \left( \frac{2}{3\epsilon^2} +\frac{10}{9\epsilon} 
-\frac{2}{3} \zeta (2)+\frac{56}{27} \right) \\
&&+ \frac{1+x^2}{(1-x)_+} \left( -\frac{2}{3\epsilon}-\frac{10}{9} \right)
+\frac{4}{3} (1+x^2) \left( \frac{\ln (1-x)}{1-x} \right)_+ -
\frac{2}{3}\frac{1+x^2}{1-x} \ln x \Bigg] \nonumber \; .
\end{eqnarray}  
The virtual--cut graph for $l^2=-l_{\perp}^2$ (corresponding to the 
gluon being an axial ghost) contributes before renormalization:
\begin{equation} \label{pp2}
\Gamma_{qq}^{({\rm g}),v} \sim \left( \frac{\alpha_s}{2\pi}\right)^2  
\delta (1-x) \; {\rm PP} \int_0^{Q^2} \frac{d|k^2| |k^2|^{-1-2 \epsilon}}
{\Gamma (1-2 \epsilon )}
\left( -\frac{2}{3\epsilon^2} -
\frac{10}{9\epsilon} -\frac{2}{3} \zeta (2)-\frac{56}{27} \right) \; .
\end{equation}  
The loop with $l^2=0$ only contributes via its renormalization 
counterterm as explained earlier. This contribution exists also for the
loop with $l^2=-l_{\perp}^2$ and reads on aggregate for both loop parts:
\begin{equation} \label{pp3}
\Gamma_{qq}^{({\rm g}),`ren'} \sim \left( \frac{\alpha_s}{2\pi}\right)^2 
{\rm PP} \int_0^{Q^2} \frac{d|k^2| |k^2|^{-1-\epsilon}}{\Gamma (1-2 \epsilon )}
\frac{2}{3} \Bigg[ \frac{1}{\epsilon} \frac{1+x^2}{(1-x)_+} -
(1+x^2) \left( \frac{\ln (1-x)}{1-x} \right)_+
-1+x \Bigg] \; .
\end{equation}  
When adding the integrands of Eqs.~(\ref{pp1})--(\ref{pp3}), all poles cancel, 
and as promised the contribution to $\Gamma_{qq}$ is finite before 
integration over $|k^2|$.

Next, we determine the contributions of the various graphs to $P_{qq}^{V,(1)}$,
making use of Eq.~(\ref{gam}). The results are displayed in Tables~1 and 2. 
We see that all entries in the tables are completely well--defined, 
even at $x=1$, in terms of distributions, which is a property that we 
already encountered at LO. 

The sums of the various graph--by--graph contributions are also presented
in Tables~1,2. One realizes that many more complicated structures, 
like the dilogarithm ${\rm Li}_2(x)$, cancel in the sums. Considering 
only $x<1$ for the moment, it is the most important finding of this work 
that the entries in the columns `Sum' in Tables~1,2 {\em exactly} reproduce 
the results found in the PV calculations~\cite{cfp,ev} for $x<1$.
Since the latter are in agreement with those obtained in the covariant--gauge
OPE calculations~\cite{ope1}, we conclude that the ML prescription has 
led to the correct final result. 
To a certain extent, this is a check on the prescription itself 
in the framework of a highly non--trivial application. Since -- in contrast 
to the PV recipe -- the ML prescription possesses a solid field--theoretical 
foundation~\cite{bas1,bas}, our calculation has finally provided 
a `clean' derivation of the NLO flavour non--singlet splitting function 
within the CFP method, highlighting the viability of that method.

The next subsection will address the endpoint ($\delta (1-x)$) contributions
to the NLO flavour non--singlet splitting function.
\begin{figure}[p]
\vspace*{-8.8cm}
\hspace*{-3.7cm}
\epsfig{file=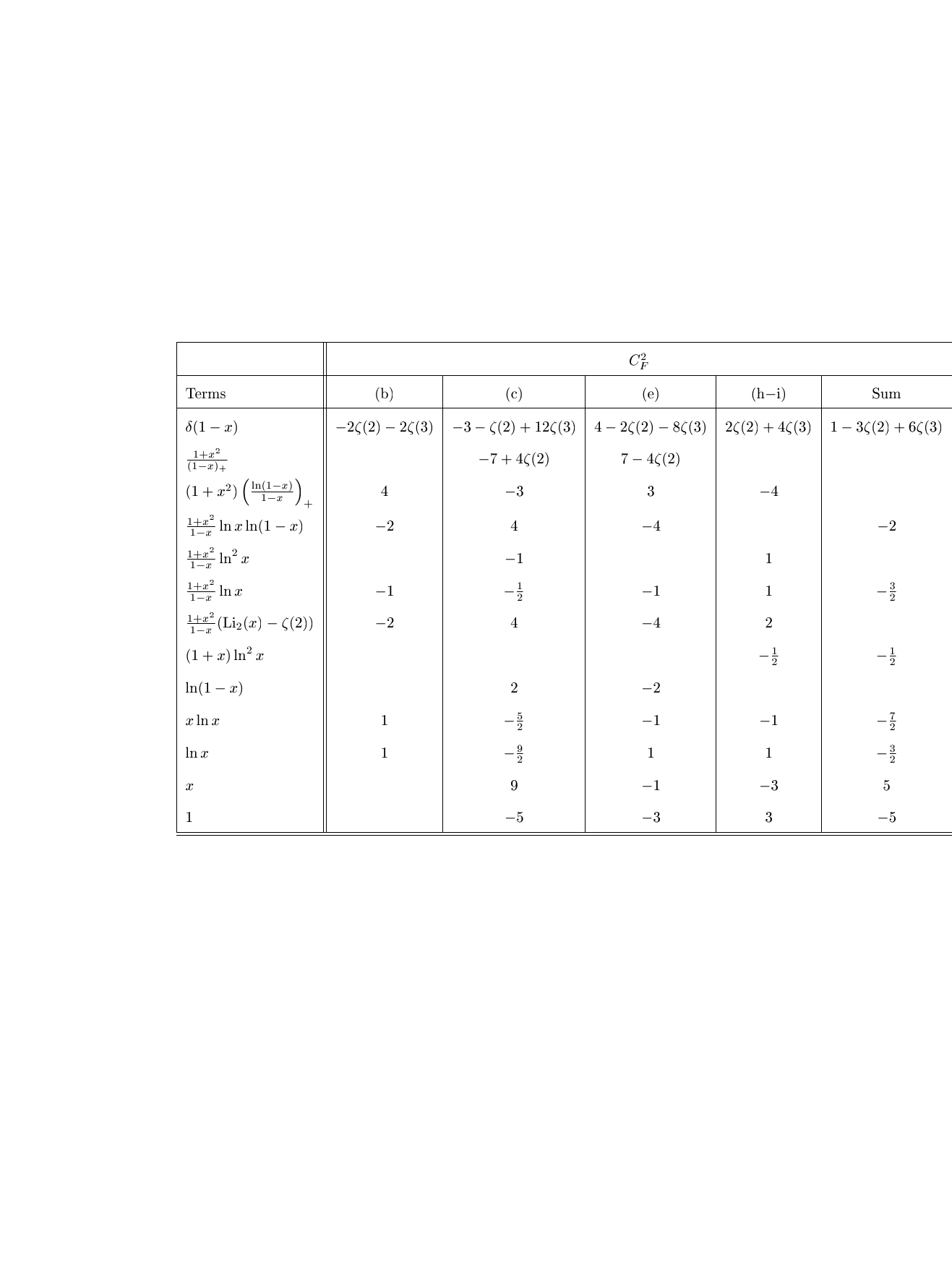,width=18.4cm}
\begin{center}
\vspace*{-9.5cm}
\end{center}
{\sf Table 1: Final results for the $C_F^2$ part of 
$P_{qq}^{V,(1)}$ on a graph--by--graph basis.}
\begin{center}
\vspace*{-7.5cm}
\end{center}
\hspace*{-4.1cm}
\epsfig{file=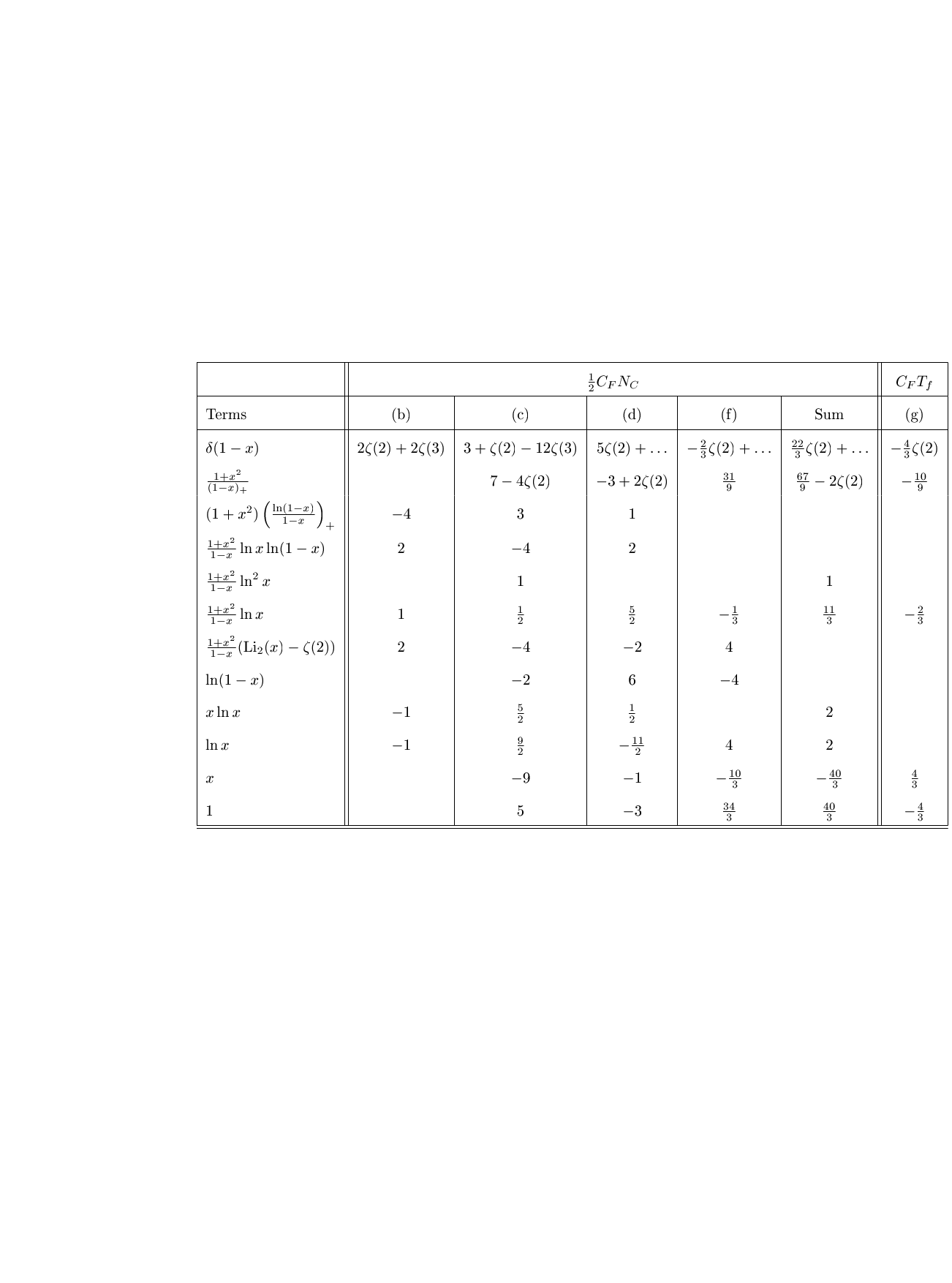,width=19cm}
\begin{center}
\vspace*{-10cm}
\end{center}
{\sf Table~2: Final results for the $C_F N_C$ and $C_F T_f$ parts of 
$P_{qq}^{V,(1)}$ on a graph--by--graph basis. The dots in the contributions
$\sim \delta (1-x)$ in the first row denote uncalculated pieces consisting of
$\zeta (3)$ and a rational number (see App.~C).}
\end{figure}
\newpage
\subsection{Contributions at $\boldmath{x=1}$ and a sample two--loop 
calculation \\ of $\boldmath{Z_q}$}
In the PV calculations~\cite{cfp,fp,ev} of the two--loop splitting
functions the contributions $\sim \delta (1-x)$ were never directly
calculated, but inferred~\cite{cfp,wada,ev} from fermion number 
conservation, expressed by the requirement
\begin{equation} \label{mom}
\int_0^1 \left( P_{qq}^{V,(1)} (x) - P_{q\bar{q}}^{V,(1)} (x) \right) dx =0 
\; .
\end{equation}
We could proceed in the same way and stop the calculation here. However,
the calculation we have performed in the ML prescription allows us
to go beyond this pragmatic approach, since -- at least for the $C_F^2$ and 
$C_F T_f$ parts -- we have always picked up the finite amounts of 
$\delta (1-x)$ contributed by the 2PI kernels. If we now performed a 
calculation of $\xi_q^{(1)}$, corresponding to the graphs in the bottom row of 
Fig.~\ref{fig2}, we would have all terms $\sim \delta (1-x)$ in the NLO 
flavour non--singlet splitting function and could check whether 
indeed~(\ref{mom}) is correctly reproduced. As an example, we will 
go this way for the $C_F T_f$ part of the splitting function.

Let us first establish what we need to get for $\xi_q^{(1)}$. The coefficient
of $\delta (1-x)$ in the NLO splitting function was determined 
in~\cite{wada,ev} via~(\ref{mom}) to be
\begin{equation} \label{ddd1}
C_F^2 \left( \frac{3}{8} - 3 \zeta (2) + 6 \zeta (3) \right) 
+C_F T_f \left( -\frac{1}{6}- \frac{4}{3} \zeta (2) \right) +
C_F N_C \left( \frac{17}{24} +\frac{11}{3} \zeta (2) - 3 \zeta (3) \right) \; ,
\end{equation}
while in our calculation we have according to Eq.~(\ref{zq}) and Tables~1,2:
\begin{equation} \label{ddd2}
\xi_q^{(1)} + C_F^2 \Bigg( 1 - 3 \zeta (2) + 6 \zeta (3) \Bigg) +
C_F T_f \left( - \frac{4}{3} \zeta (2) \right) + 
C_F N_C \left( \frac{11}{3} \zeta (2) + \ldots \right) \; ,
\end{equation}
where the dots indicate that we have not entirely calculated the finite 
amount of $\delta (1-x)$ in the $C_F N_C$ part of $P_{qq}^{V,(1)}$, 
even though we were able to determine its contribution $\sim \zeta (2)$
(see App.~C). Comparing Eqs.~(\ref{ddd1}) and (\ref{ddd2}), we get a
prediction for the $C_F^2$ and $C_F T_f$ parts of $\xi_q^{(1)}$ in the 
light--cone gauge with ML prescription:
\begin{equation} \label{xipr}
\xi_q^{(1)} = - \frac{5}{8} C_F^2 -\frac{1}{6} C_F T_f +C_F N_C \Big( 
\ldots \Big) \; ,
\end{equation}
where all we can say about the $C_F N_C$ part is that it does not contain
any terms $\sim \zeta (2)$. It is quite remarkable that no $\zeta (2)$, 
$\zeta (3)$ terms are left over in the $C_F^2$ and $C_F T_f$ parts of 
$\xi_q^{(1)}$.

To directly calculate the $C_F T_f$ part of $\xi_q^{(1)}$, we only have 
to evaluate the first diagram in the bottom row of Fig.~\ref{fig2}. What we 
need to extract is the two--loop renormalization constant for that
diagram, when the light--cone gauge with the ML prescription is used.
The calculation is relatively easy since the inner quark loop has obviously 
no light--cone gauge propagator and can in fact be calculated exactly:
\begin{equation} \label{pi}
\Pi_{\mu\nu}(r) = -i T_f \frac{\alpha_s}{4\pi} 
\frac{8 \Gamma^2 (2-\epsilon) \Gamma(\epsilon)}{\Gamma (4-2 \epsilon)}
\left( \frac{4 \pi}{-r^2} \right)^{\epsilon} 
\Big[ r^2 g_{\mu\nu}-r_{\mu} r_{\nu} \Big] \; .
\end{equation}
This self--energy can then be renormalized with the help of the 
counterterm in Fig.~\ref{fig3}. The renormalized loop is then inserted 
into the outer loop. Here it is very convenient that $\Pi_{\mu\nu}$ is
transverse, 
\begin{equation}
{\cal D}^{\alpha\mu} (r) \Big[ r^2 g_{\mu\nu}-r_{\mu} r_{\nu} \Big]
{\cal D}^{\nu\beta} (r) \sim {\cal D}^{\alpha\beta} (r) \; .
\end{equation}
In other words, the whole calculation is not very different from a simple
one--loop calculation of the quark self--energy, the only exception being 
that we now need loop integrals with the extra factor $\left( -r^2
\right)^{-\epsilon}$ present in~(\ref{pi}). If we embed the whole 
graph into the Dirac trace as shown in Fig.~\ref{fig1}(b), it turns out that 
we only need few integrals of this kind; they are collected in Appendix~E. 
Since we have renormalized the inner loop, the left--over divergence 
after loop--integration determines the two--loop counterterm and thus the
contribution to $\xi_q^{(1)}$. We find in the $\overline{\rm{MS}}$ scheme:
\begin{equation}
Z_q^{C_F T_f} = 1+ \left( \frac{\alpha_s}{2\pi}\right)^2 C_F T_f 
\left( - \frac{1}{2\epsilon^2} + \frac{1}{12\epsilon} \right) \; .
\end{equation}
Comparing to Eq.~(\ref{zq}) this implies that the $C_F T_f$ part 
of $\xi_q^{(1)}$ is exactly what we expected it to be in~(\ref{xipr}):
\begin{equation}
\xi_{q}^{(1),C_F T_f} = -\frac{1}{6} C_F T_f \; .
\end{equation}
This result clearly demonstrates the consistency of the whole approach: our
example shows that the light--cone gauge method of~\cite{egmpr,cfp,fp} 
is also able to determine the contributions $\sim \delta (1-x)$ to the 
splitting functions by explicit calculation. It would be interesting in
this context to calculate also the other contributions to $\xi_q^{(1)}$ 
(and the missing part $\sim C_F N_C \delta (1-x)$ in our Tab.~2); important 
steps in this direction have been taken in~\cite{lw,wthes} by examining the 
second and the third diagram in the bottom row of Fig.~\ref{fig2}, which 
yield the $C_F^2$ part of $\xi_q^{(1)}$. Indeed it turns out that the results 
of~\cite{lw,wthes} can be exploited to reproduce the term $-5C_F^2/8$
in our prediction~(\ref{xipr}) for $\xi_q^{(1)}$, which can be regarded as
a further confirmation of our results. 

We have to admit, however, at this point that the ability to obtain the correct
endpoint contributions is not restricted to the ML prescription: this is 
also possible for the PV prescription. Here, the coefficient of 
$\delta (1-x)$ in the $C_F T_f$ part of $P_{qq}^{V,(1)}$ reads
\begin{equation} \label{pvx1}
\xi_{q,{\rm PV}}^{(1),C_F T_f} - C_F T_f \frac{20}{9} I_0 \; .
\end{equation}
Here the second term originates from the entry `$-10/9$' in the last column
of Tab.~2, when we omit the `plus'--prescription there and reintroduce it
using the PV identity~\cite{cfp}
\begin{equation}
\frac{1}{1-x} \longrightarrow I_0 \delta (1-x) + \frac{1}{(1-x)_+} \; ,
\end{equation}
where $I_0$ is as defined in~(\ref{io}). Furthermore, 
$\xi_{q,{\rm PV}}^{(1),C_F T_f}$ denotes the $C_F T_f$ part of $\xi_q^{(1)}$ 
when the PV prescription is used. The explicit calculation gives 
\begin{equation}
Z_{q,{\rm PV}}^{C_F T_f} = 1+ \left( \frac{\alpha_s}{2\pi}\right)^2 C_F T_f 
\Bigg[ - \frac{1}{2\epsilon^2} \left( 1-\frac{4}{3} I_0 \right) + 
\frac{1}{\epsilon} \left( \frac{1}{12} + \frac{2}{3} \zeta (2) - 
\frac{10}{9} I_0
\right) \Bigg] \; ,
\end{equation}
that is 
\begin{equation} \label{pvx2}
\xi_{q,{\rm PV}}^{(1),C_F T_f} = C_F T_f \left( -\frac{1}{6} - \frac{4}{3} 
\zeta (2) +\frac{20}{9} I_0 \right) \; .
\end{equation}
It is interesting to see how upon combining Eqs.~(\ref{pvx1}) and (\ref{pvx2})
the $I_0$ terms drop out, and the $C_F T_f$ part of the endpoint 
contributions comes out correctly as in~(\ref{ddd1})
also for the PV prescription. We note, however, that again this happens at the
expense of having renormalization constants depending on singular quantities 
like $I_0$ that represent a mix--up in the treatment of UV and 
IR singularities.
\newpage
\section{The calculation of the ${\boldmath N_C^2}$ part of the
singlet \\ splitting function ${\boldmath P_{gg}}$ at NLO}
Let us now turn to the calculation of $P_{gg}^{(1)}$. We restrict 
ourselves to its $N_C^2$ part, since the contributions $\sim C_F T_f$, 
$N_C T_f$ are essentially trivial as far as the treatment of the LCA is 
concerned: The $C_F T_f$ part comprises no gluon emission at all, and all 
diagrams contributing to the $N_C T_f$ part contain a quark loop and the 
emission of at most {\em one} gluon. Such diagrams with one--gluon emission 
have the LO kinematics and will not reveal any new features as compared
to what we have already discussed. In contrast to this, the $N_C^2$ part
of $P_{gg}^{(1)}$ requires the renormalization of the non--Abelian part of the 
three--gluon vertex and therefore really provides a further challenge for
the ML prescription.

The diagrams contributing to the $N_C^2$ part of $\Gamma_{gg}$ at NLO are 
shown in Fig.~4. We do not show here the graphs contributing to $Z_g$ at 
two loops, since we will not attempt to calculate them. 
\begin{figure}[b]
\begin{center}
\vspace*{-13cm}
\end{center}
\hspace*{0.2cm}
\epsfysize23cm
\leavevmode\epsffile{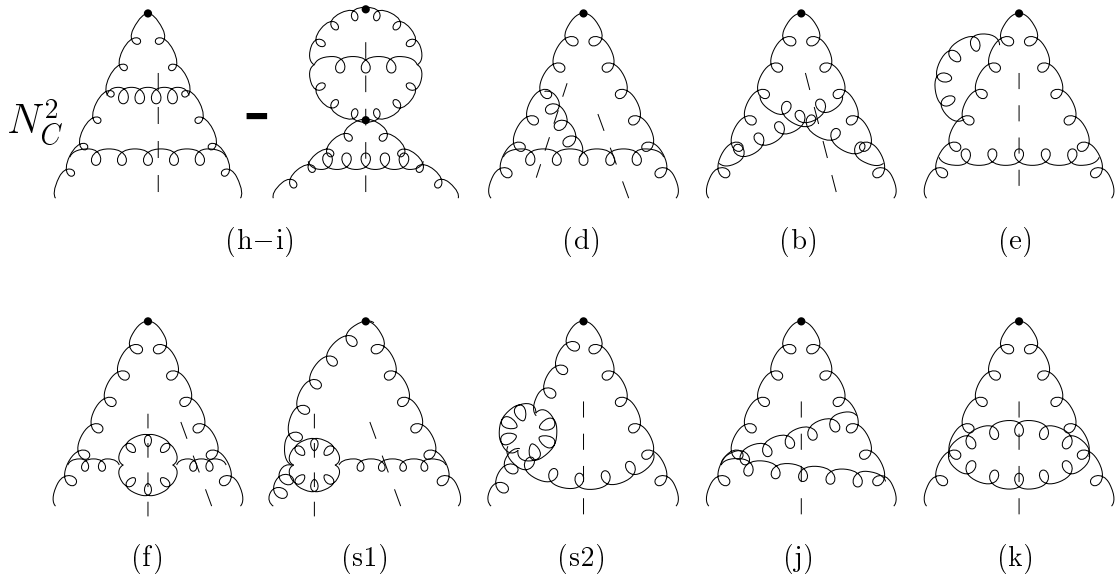}
\begin{center}
\vspace*{-13.5cm}
\caption{\label{fig4}{\sf Diagrams contributing to the $N_C^2$ part of 
$\Gamma_{gg}$ at NLO.}}
\end{center}
\end{figure}
\newpage

The calculation of the various real--cut and virtual--cut diagrams proceeds
in exactly the same way as before. For the renormalization of the 
virtual--cut contributions in the triangle graph (d) and the `swordfish' 
ones (s1),(s2), we need the UV counter--term for the three--gluon 
vertex in the light--cone gauge with ML prescription. Here we can rely 
on the result presented in~\cite{dal3g} (see also~\cite{leibbook,basbook}); 
the part of it that is relevant for our calculation is recalled in Fig.~5.

Concerning the real cuts, we mention that for graphs (h),(b),(j),(k)
we use the expression in~(\ref{b5}) for the phase space. As for the $C_F N_C$ 
part of $P_{qq}^{V,(1)}$, we found it easier to determine the contributions 
\begin{figure}[hb]
\begin{center}
\hspace*{-0.6cm}
\vspace*{0.5cm}
\epsfysize24cm
\leavevmode\epsffile{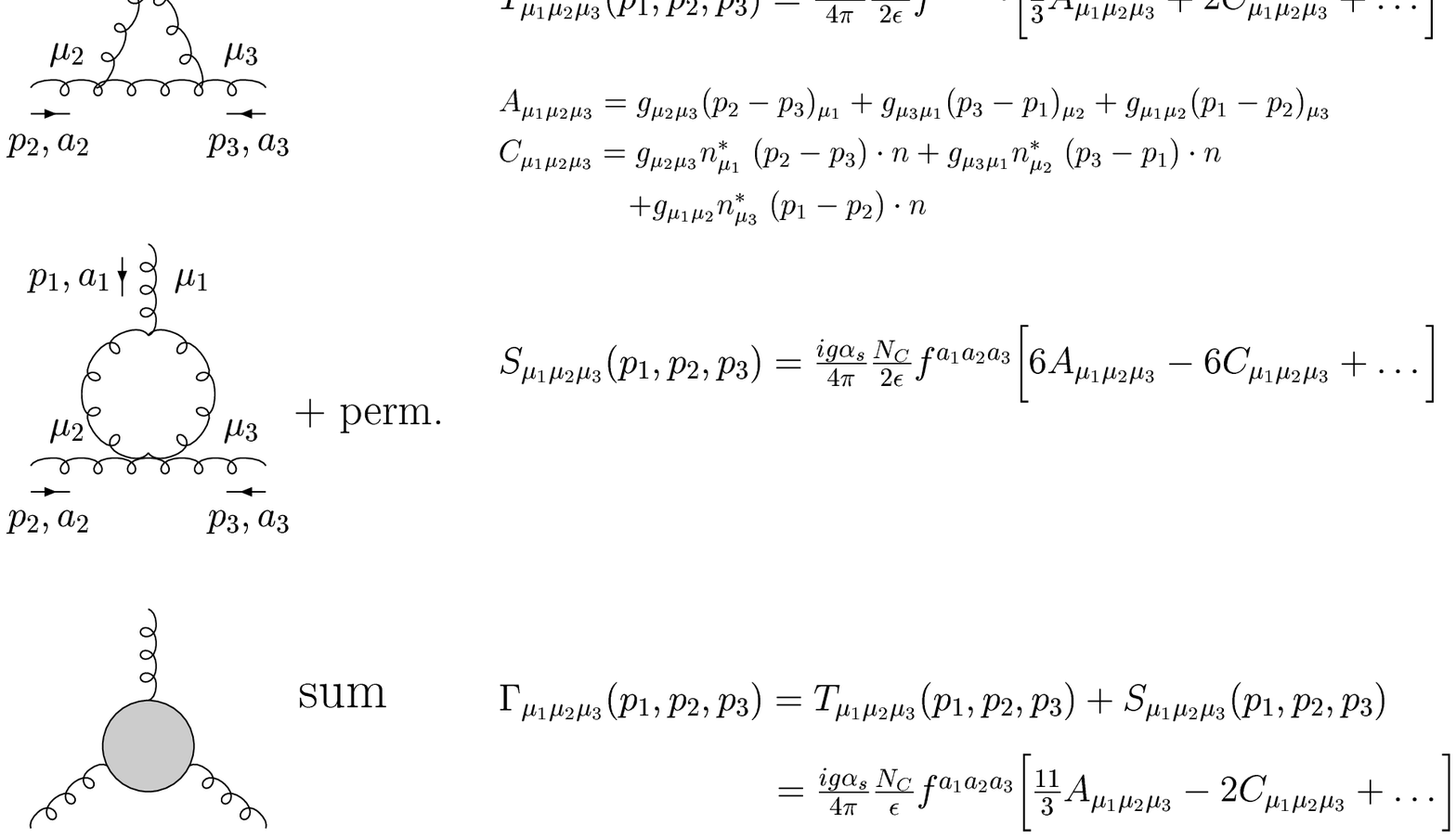}
\end{center}
\vspace*{-14cm}
\caption{\label{fig5}{\sf UV--divergent structures of the non--Abelian
part of the three--gluon vertex as obtained in the light--cone gauge, 
using the ML prescription. $p_1$,$p_2$,$p_3$ denote the momenta of the 
external gluons, $a_1$,$a_2$,$a_3$ are the associated colour indices
($f^{a_1 a_2 a_3}$ being the structure constants of SU(3)), and 
$\mu_1$,$\mu_2$,$\mu_3$ are Lorentz indices. The dots indicate structures 
(some of them non--local) which do not contribute to our calculation 
thanks to the orthogonality of the free propagator to the gauge vector $n$.}}
\end{figure} 
of the real cuts of the remaining diagrams via the
extraction of the imaginary parts of the associated virtual graphs. 

We have verified that again for the ML prescription all 2PI graphs
give truly finite contributions to $\Gamma_{gg}$, before the final 
integration over $|k^2|$ is performed. This also applies to the endpoint
$x=1$, where the result for each graph is again always well--defined 
in terms of distributions and, as before, also has a coefficient of 
$\delta (1-x)$ that contains no $1/\epsilon$ poles. Table~3 presents the 
contributions of the various diagrams to $P_{gg}^{(1)}$. Here we have defined
the functions 
\begin{eqnarray}
p_{gg}(x) &\equiv& \frac{\left( 1-x+x^2 \right)^2}{x (1-x)_+} \; , \nonumber 
\\
l_{gg}(x) &\equiv& \frac{\left( 1-x+x^2 \right)^2}{x} \Bigg( 
\frac{\ln (1-x)}{1-x} \Bigg)_+ \; , \\
S_2 (x) &\equiv& \int_{\frac{x}{1+x}}^{\frac{1}{1+x}} \frac{dz}{z} 
\ln \Big(\frac{1-z}{z} \Big) = -2 {\rm Li}_2 (-x)-2 \ln x \ln (1+x)+
\frac{1}{2} \ln^2 x- \zeta(2) \; .  \nonumber 
\end{eqnarray}
We mention in passing that graph (j) and the `swordfish' diagram (s1) give 
vanishing contributions to $P_{gg}^{(1)}$ if the PV prescription is used, 
but are non--vanishing for the ML prescription, where finite contributions
arise from their ghost parts.

As for the case of $P_{qq}^{V,(1)}$, the full result for the $N_C^2$ 
part of $P_{gg}^{(1)}$, given by the column `Sum', is (at $x<1$) in 
agreement with the PV result of~\cite{fp}, which in turn coincides
with the OPE\footnote{See also our discussion in the introduction concerning 
the OPE calculations~\cite{hvn,ope2} of $P_{gg}^{(1)}$.} 
one~\cite{hvn}. Thus, the CFP method with ML prescription 
has also led to the correct final answer in this case, which clearly 
constitutes a further non--trivial and complementary check. As can be seen 
from Tab.~3, we have not determined the finite amounts of contributions
$\sim \delta (1-x)$ for the graphs since, like in the case of the 
$C_F N_C$ part of $P_{qq}^{V,(1)}$, these are quite hard to extract in some
cases. The endpoint contributions to $P_{gg}^{(1)}$ can then only be derived
from the energy--momentum conservation condition~\cite{wada,ev}. We 
emphasize however that, just as for $P_{qq}^{V,(1)}$, there is no principal
problem concerning the calculation of the endpoint contributions: had we 
calculated the full $\delta (1-x)$--terms in Tab.~3 and the two--loop
quantity $\xi_g^{(1)}$, all endpoint contributions would be at our disposal, 
and it would no longer be necessary to invoke the energy--momentum 
conservation condition; in fact, this could serve as a further check of
the calculation. 
\begin{figure}[p]
\vspace*{-4.8cm}
\hspace*{-4.4cm}
\epsfig{file=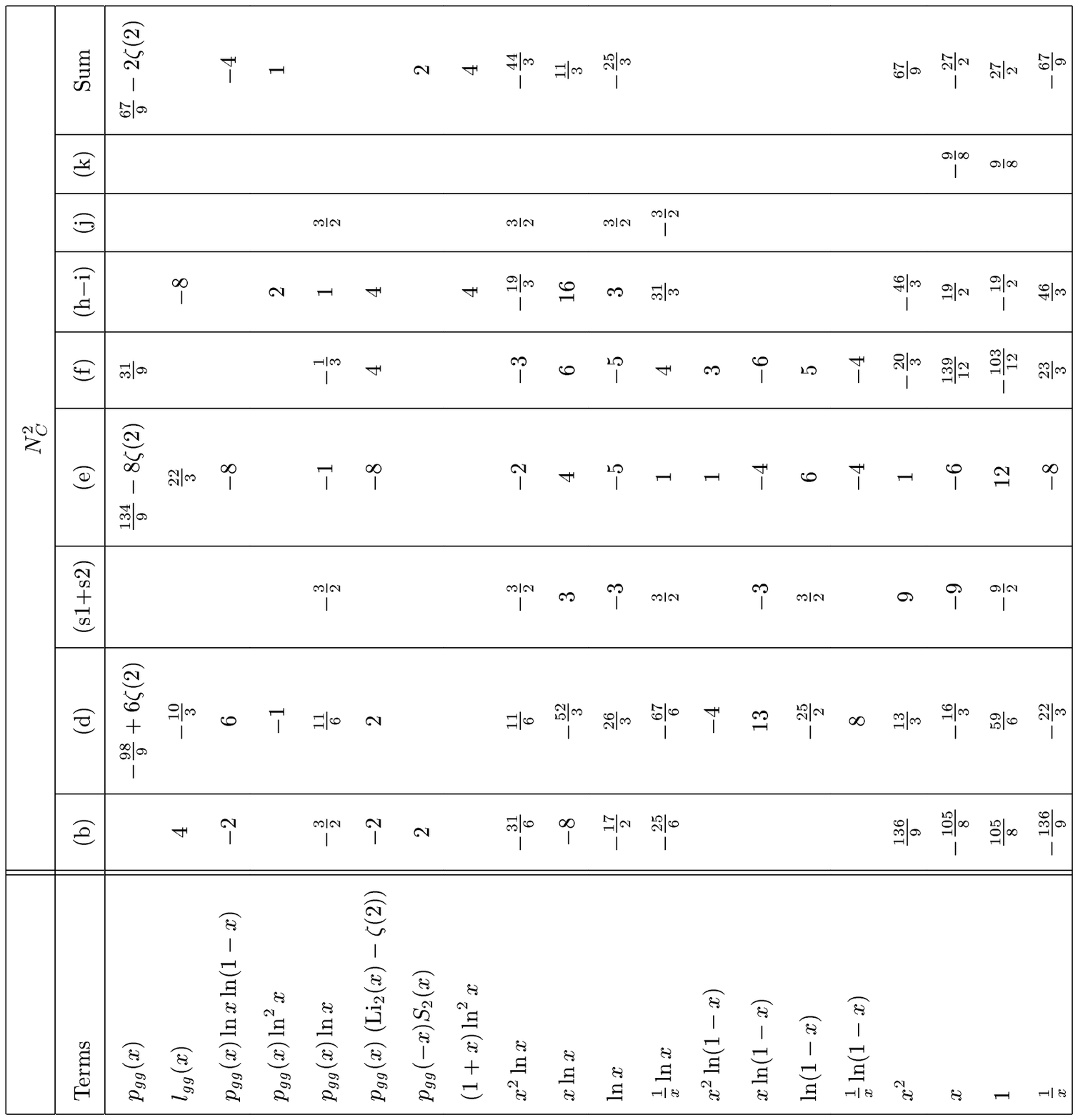,width=19.8cm}
\begin{center}
\vspace*{-6.3cm}
\end{center}
{\sf Table 3: Final results for the $N_C^2$ part of 
$P_{gg}^{(1)}$ on a graph--by--graph basis. The table does not include
the coefficients of $\delta (1-x)$, which we have not determined. However, 
as mentioned in the main text, we have proven that each graph 
contributes a {\em finite} amount of $\delta (1-x)$ to $\Gamma_{gg}$ 
(before the final integration over $|k^2|$ is performed).}
\end{figure}
\section{Conclusions}
We have performed a new evaluation of the NLO flavour non--singlet 
splitting function and of the $N_C^2$ part of the NLO gluon--to--gluon 
splitting function within the light--cone gauge method of~\cite{egmpr,cfp}.
The new feature of our calculation is the use of  
the Mandelstam--Leibbrandt prescription for dealing with the 
spurious poles generated by the gauge denominator in the gluon propagator.
In contrast to the principal value prescription employed in previous 
calculations~\cite{cfp,fp,ev}, the ML prescription has a solid 
field--theoretical foundation and will therefore provide a `cleaner'
derivation of the result. As expected, the final results come out correctly,
i.e., are in agreement with the ones in~\cite{cfp,fp,ev,ope1}. This finding
is both a corroboration of the usefulness of the general method 
of~\cite{egmpr,cfp} to calculate splitting functions, {\em and} a useful
check on the ML prescription itself in a highly non--trivial application.

We have also discussed the $\delta (1-x)$ contributions to the NLO 
flavour non--singlet splitting 
function, performing an explicit sample calculation of a two--loop 
contribution to the renormalization constant $Z_q$ in the ML 
light--cone gauge. It turns out that one indeed obtains the right amount 
of contributions at $x=1$ as required by fermion number conservation. 

We conclude by conceding that the ML prescription is in general much more
complicated to handle than the simpler, but less well--founded, PV 
prescription. With regard to future applications at, for instance, three--loop
order, this creates a certain dilemma: the ML prescription might
be too complicated to be used in that case, while on the other hand
the ill--understood success of the PV prescription at the two--loop level
is not a warranty that it will also produce correct results beyond.
\vspace*{-0.3cm}
\section*{Acknowledgement}
\vspace*{-0.4cm}
This work was supported in part by the EU Fourth Framework Programme 
`Training and Mobility of Researchers', Network `Quantum Chromodynamics and 
the Deep Structure of Elementary Particles', contract FMRX-CT98-0194 
(DG 12-MIHT). 
\newpage
\section*{Appendix A: Virtual integrals}
\appendix
\setcounter{equation}{0}
\renewcommand{\theequation}{A.\arabic{equation}}
Here we list some loop integrals needed for the calculation. We do not
need to recall any of the covariant integrals, which are standard, 
but will only present those with a light--cone gauge denominator, to be 
treated according to the ML prescription~(\ref{propml}).

We begin by performing a sample calculation of the integral 
\begin{eqnarray}
I (n,q) &\equiv& \int \frac{d^d r}{(2\pi)^d} \frac{1}{(r^2+i \varepsilon) 
( (q-r)^2+i \varepsilon) [nr]} \nonumber \\
&=& \int \frac{d^d r}{(2\pi)^d}  \frac{n^*r}{(r^2+i \varepsilon) 
( (q-r)^2+i \varepsilon) (nr \, n^*r+i \varepsilon)} \; .
\end{eqnarray}
We recall the definitions~\cite{cfp}
\begin{equation}
n=\frac{pn}{2P} (1,0,\ldots,0,-1) \; , \;\;\; n^*=\frac{P}{pn} 
(1,0,\ldots,0,1) \equiv \frac{1}{pn} p \; ,
\end{equation}
where $p=P(1,0,\ldots,0,1)$ is the momentum of the 
incoming quark, see Fig.~\ref{fig2}. Introducing Feynman parameters, one has
\begin{equation}
I (n,q)=\frac{4P}{pn} \int_0^1 dt \int_0^{1-t} ds \int \frac{d^d r}{(2\pi)^d}
\frac{r_0-r_z}{\Big[ r^2 +s r_{\perp}^2-2 (q\cdot r) t+q^2 t 
+i \varepsilon \Big]^3} \; .
\end{equation}
After performing a Wick rotation and straightforward integrations over $r$
one arrives at 
\begin{equation} \label{a4}
I (n,q)=\frac{i\Gamma(1+\epsilon)}{16\pi^2} \frac{2 n^* q}{q^2+i \varepsilon} 
\Big( \frac{4 \pi}{-q^2} \Big)^{\epsilon} 
\int_0^1 dt \, t^{-\epsilon} (1-t)^{-1-\epsilon} \int_0^1 ds
\Bigg( 1+s t \frac{q^2+q_{\perp}^2+i \varepsilon}{(q^2+i \varepsilon) 
(1-t)} \Bigg)^{-1-\epsilon} .
\end{equation}
For example, for the case $q=k$ one finds
\begin{equation} \label{a5}
I (n,k)=\frac{i\Gamma(1+\epsilon)}{16\pi^2} \Big( \frac{4 \pi}{|k^2|} 
\Big)^{\epsilon} \frac{1}{[nk]} \Bigg[ \zeta(2) - {\rm Li}_2 
\left( \frac{k_{\perp}^2}{|k^2|} \right) +2 \epsilon \zeta(3) \Bigg] \; ,
\end{equation}
where we have kept those terms that contribute to the final answer.
In~(\ref{a5}), $\zeta(n)$ is Riemann's $\zeta$--function and ${\rm Li}_2 (x)$
denotes the dilogarithm, defined by~\cite{dd}
\begin{equation}
{\rm Li}_2 (z) \equiv - \int_0^1 \frac{\ln (1-z y)}{y} dy \; .
\end{equation}
The result in~(\ref{a5}) coincides with the one
in~\cite{basbook} for $\epsilon=0$. Note that the ML prescription arising for 
$1/[nk]$ is actually immaterial here since $nk=x \, pn$ never vanishes.

Setting, on the other hand, $q=l$ one gets for $l^2=0$
\begin{equation} \label{a6}
I (n,l) = \frac{i\Gamma(1+\epsilon)}{16\pi^2} \Big( \frac{4 \pi}
{-|k^2| (1-x)} \Big)^{\epsilon} \frac{1}{nl} \; \frac{1}{2\epsilon^2} 
\; ,
\end{equation} 
and for $l^2=-l_{\perp}^2$ 
\begin{equation}
I (n,l) =\frac{i\Gamma(1+\epsilon)}{16\pi^2} 
\Big( \frac{4 \pi}{-l^2} \Big)^{\epsilon} \; \frac{2 n^* l}{l^2} \; 
B (-\epsilon,1-\epsilon) \; . 
\end{equation} 
Note that the real part of~(\ref{a6}) has to be taken\footnote{Here one
obviously has to discard the overall factor $i$.}. 
Table~4 contains all the required integrals with an ML light--cone gauge 
denominator. The integrals in the first column are for $l^2=0$; they 
depend on 
\begin{equation} \label{xtdef}
x = \frac{nk}{pn} \;\;\; \;\; {\rm and} \;\; \;\;\; 
\tilde{x}=\frac{nl}{pn} = 1-x \; . 
\end{equation}
Recall that terms $\sim (1-x)^{-1-a \epsilon}$ will lead to further poles, 
as was shown by~(\ref{del}). The integrals in the second column of Tab.~4 
are for the axial ghost case, $l^2=-l_{\perp}^2$, and eventually need to 
be integrated further over the variable $\kappa$ defined in 
Eqs.~(\ref{ps2}),(\ref{def2}). The $\kappa$ integration produces further 
poles. We have 
\begin{equation} \label{ktdef}
\kappa = \frac{k_{\perp}^2}{|k^2|} \;\;\; \;\; {\rm and} \;\; \;\;\; 
\tilde{\kappa}=1-\kappa \; . 
\end{equation}
We note that the last integral was much easier obtained by performing the 
$\kappa$ integration before the ones over the Feynman parameters. 
Therefore we only present the final, $\kappa$--integrated, result in this 
case. As can be seen, the integral was accompanied by two different powers
of $\kappa$.  
\newpage
\setcounter{table}{3}
\begin{table}[ht]
\begin{center}
\renewcommand{\arraystretch}{1.6}
\begin{tabular}{|c|c|c|} \hline \hline
$\int d^d r/(2\pi)^d$ & $l^2=0$ & $l^2+l_{\perp}^2=0$  \\ \hline \hline
$\frac{pn}{r^2 (k-r)^2 [nr]}$ & $\frac{1}{x} \Big[ \zeta(2) - {\rm Li}_2 
\left( \tilde{x} \right) +2 \epsilon \zeta(3) \Big]$ &
$\zeta(2) - {\rm Li}_2 \left( \kappa \right) +2 \epsilon \zeta(3)$ \\ \hline
$\frac{(r^2-2 nk \; n^* r) \; pn}{r^2 (k-r)^2 [nr]}$ & 
$k^2 \Big[ \frac{1}{\epsilon} +2 (1-\ln x) +
\frac{1}{x} {\rm Li}_2 (\tilde{x})$ &
$k^2 \tilde{\kappa} \Big[ \frac{1}{\epsilon} +
2 (1-\ln \tilde{\kappa} ) +\frac{1}{\tilde{\kappa}} {\rm Li}_2 
(\kappa)$ \\ 
& $-\frac{1}{x}\zeta(2)+\epsilon \left (4+\zeta (2) -2 \zeta(3) \right) 
\Big] $ & $-\frac{1}{\tilde{\kappa}} \zeta(2)
+\epsilon \left (4+\zeta (2) -2 \zeta(3) \right) \Big] $
\\ \hline
$\frac{pn}{r^2 (l-r)^2 [nr]}$ & $\frac{1}{2\epsilon^2} 
\tilde{x}^{-1-\epsilon} (1-\epsilon^2 \zeta (2)+2 \epsilon^3 \zeta (3))$ & 
$-\kappa^{-\epsilon} \frac{1}{\epsilon}
\left( 1-\frac{1}{\kappa} \right) (1+\epsilon^2 \zeta (2))$ \\ \hline 
$\frac{(r^2-2 nl \; n^* r)\; pn}{r^2 (l-r)^2 [nr]}$ &
$-2 k^2 \tilde{x}^{-\epsilon} 
\Big[ \frac{1}{\epsilon} +2 +\epsilon (4-\zeta(2) )
\Big] $ & 0 \\ \hline
$\frac{pn}{(p-r)^2 (k-r)^2 [nr]}$ & $\frac{1}{\tilde{x}} \Big[
\frac{\ln x}{\epsilon} -\frac{1}{2} \ln^2 x -{\rm Li}_2 (\tilde{x})$ & 
$\left(\frac{1}{\epsilon}+2+4 \epsilon +
\epsilon \zeta (2) \right) \left( \kappa^{-\epsilon} -1 \right)$
\\&$ -\epsilon \tilde{x} +2 \epsilon \zeta (2) \ln x \Big]$ & 
$+ \left( 1-\frac{1}{\kappa} \right) \ln \tilde{\kappa} +\epsilon
(3-\zeta(2))$ \\ \hline
$\frac{pn}{(p-r)^2 (l-r)^2 [nr]}$ & does not occur & 
$\frac{1}{\epsilon^2} (1+\epsilon^2 \zeta (2)) 
\left( \kappa^{-\epsilon} -1 \right)$ \\ \hline
$\frac{pn}{(k+r)^2 (l-r)^2 [nr]}$ & 
$\left( \tilde{x}^{-\epsilon} - x^{-\epsilon} \right)
\left(\frac{1}{\epsilon^2}-\zeta (2)+2 \epsilon \zeta (3) \right)$ & 
$-\tilde{\kappa}^{-\epsilon} \Big[ \frac{1}{\epsilon^2} 
+2 \zeta (2) + 2 \epsilon \zeta (3) \Big]$ \\
&$-3 \zeta(2) x^{-\epsilon}$ & \\ \hline 
$\frac{pn}{r^2 (p-r)^2 (k-r)^2 [nr]}$ & $\frac{1}{k^2} \Big[
\frac{1}{\epsilon^2} +\frac{\ln x}{\epsilon} -2 {\rm Li}_2 (\tilde{x})$ & 
$\frac{1}{k^2} \Big[ \frac{1}{\epsilon^2} \left( 1- \kappa^{-\epsilon} 
\right)$ \\ 
&$-\frac{1}{2} \ln^2 x - 2 \epsilon \zeta (3) \Big] $ & 
$-\kappa^{-\epsilon} \left( {\rm Li}_2 (\kappa) + \zeta(2) \right)$ \\
& & $-2 \left( {\rm Li}_2 (\tilde{\kappa}) -\zeta (2)+\epsilon \zeta (3) 
\right) \Big] $ \\ \hline
$\frac{pn}{r^2 (k+r)^2 (l-r)^2 [nr]}$ & $
\frac{\tilde{x}^{-1-\epsilon}}{k^2}
\Big[ \frac{3}{2\epsilon^2} +\frac{\ln x}{\epsilon}-{\rm Li}_2 (x)$ & 
$\frac{1}{k^2 \kappa} \Big[ \frac{1}{\epsilon} \ln \tilde{\kappa}
+{\rm Li}_2 (\kappa)-\frac{1}{2} \ln^2 \tilde{\kappa}$ \\
&$-\frac{1}{2} \ln^2 x-\frac{5}{2} \zeta (2)-\epsilon
\left( \tilde{x}+5 \zeta (3) \right)$ & 
$-\kappa^{-\epsilon} \left( \frac{1}{\epsilon^2} +3 \zeta (2)+6 \epsilon
\zeta (3) \right) \Big]$ \\
&$-\frac{1}{2} \epsilon \ln \tilde{x} ( \ln^2 x -2 {\rm Li}_2 (\tilde{x})
-\ln x \ln \tilde{x})\Big]$ & \\ \hline
$\frac{pn}{r^2 (p-r)^2 (l-r)^2 [nr]}$ & does not occur & 
$\frac{1}{k^2} \Big[ \frac{1}{2\epsilon^3}-\frac{\zeta(2)}{2\epsilon}
-3 \zeta(3) \Big]$ \\ & & (after integration $\int_0^1 d\kappa 
\kappa^{-\epsilon}$) \\
& & $\frac{2}{k^2} \Big[ \frac{1}{\epsilon} +3+\zeta (2) \Big] $ \\
& &  (after integration $\int_0^1 d\kappa \kappa^{1-\epsilon}$) \\ \hline
\end{tabular}
\vspace*{0.3cm}
\caption{\sf Two-- and three--point integrals with a light--cone gauge
denominator for the ML prescription, calculated up to 
${\cal O}(\epsilon)$. We have dropped the ubiquitous factor 
$i/16\pi^2 \left( 4\pi/|k^2| \right)^{\epsilon} \Gamma(1-\epsilon)/
\Gamma(1-2 \epsilon)$. $\tilde{x}$ and $\tilde{\kappa}$ have been defined
in Eqs.~(\ref{xtdef}) and (\ref{ktdef}), respectively.}
\vspace*{-0.5cm}
\end{center}
\end{table}
\section*{Appendix B: Three--particle phase space}
\appendix
\setcounter{equation}{0}
\renewcommand{\theequation}{B.\arabic{equation}}
As we discussed in Sec.~1, the phase space for two gluons (plus one
`observed' parton) will split up into four pieces for the ML prescription:
\begin{eqnarray} \label{b1}
PS_1 &=& x\int d^d l_1 d^d l_2 \,\delta \left(1- x-\frac{nl_1+nl_2}{pn} \right)
\delta (l_1^2) \delta (l_2^2) \; , \nonumber \\
PS_2 &=& x\int d^d l_1 d^d l_2 \,\delta \left(1- x-\frac{nl_1+nl_2}{pn} \right)
\delta (l_1^2+l_{1,\perp}^2) \delta (l_2^2) \; , \nonumber \\
PS_3 &=& x\int d^d l_1 d^d l_2 \,\delta \left(1- x-\frac{nl_1+nl_2}{pn} \right)
\delta (l_1^2) \delta (l_2^2+l_{2,\perp}^2) \; , \nonumber \\
PS_4 &=& x\int d^d l_1 d^d l_2 \,\delta \left(1- x-\frac{nl_1+nl_2}{pn} \right)
\delta (l_1^2+l_{1,\perp}^2) \delta (l_2^2+l_{2,\perp}^2) \; ,
\end{eqnarray}
where $l_1$,$l_2$ are the gluon momenta. The $\delta$ functions
in~(\ref{b1}) determine whether one (or both) of the gluons acts as an axial 
ghost.

As we know from the discontinuity in~(\ref{disc}), the tensorial 
structures of the non--ghost part and the ghost part are different.
However, we can rewrite~(\ref{disc}) as
\begin{equation} \label{disc1}
\mbox{{\rm disc}}\left[{\cal D}^{\mu\nu} (l)\right] =
2 \pi \Theta(l_0) \frac{2 n^* l}{l_{\perp}^2} \Bigg( 
\delta (l^2) - \delta (l^2+l_{\perp}^2) \Bigg) 
\Bigg[ -g^{\mu \nu} (nl) + n^\mu l^\nu+n^\nu l^\mu \Bigg] \; .
\end{equation}  
This is possible because of $2 n^* l/l_{\perp}^2=1/nl$ for $l^2=0$ and
$(nl)(n^*l)=0$ for $l^2+l_{\perp}^2=0$. In this way, it is always possible to
calculate just one combined matrix element, using the tensorial structure in
square brackets, and integrate it over a phase space subject to 
simply the difference $\delta (l^2) - \delta (l^2+l_{\perp}^2)$. 
For our two--gluon case, this means that we have to consider only the 
combination
\begin{equation} \label{b3}
PS_1-PS_2-PS_3+PS_4 \; . 
\end{equation}
We now introduce the Sudakov parametrizations
\begin{eqnarray}
l_1^{\mu} &=& (1-z) p^{\mu} + \frac{l_1 p}{pn} n^{\mu} + l_{1,\perp}^{\mu} \; ,
\nonumber \\
l_2^{\mu} &=& z(1-y) p^{\mu} + \frac{l_2 p}{pn} n^{\mu} + l_{2,\perp}^{\mu} 
\; ,
\end{eqnarray}
where $(l_{i,\perp}^{\mu})^2=-l_{i,\perp}^2$. The first
$\delta$ functions in~(\ref{b1}) imply $y=x/z$. 
If one wants to integrate over an arbitrary function $f$ of scalar products
of the momenta, one writes the four parts of phase space in the following way:
\begin{eqnarray} \label{b5}
\lefteqn{
x\int d^d l_1 d^d l_2 \,\delta \left(1- x-\frac{nl_1+nl_2}{pn} \right)
f\Big( \frac{l_{1,\perp}^2}{|k^2|},\frac{-(p-l_1)^2}{|k^2|},
\frac{l_{2,\perp}^2}{|k^2|},\frac{l_2 p}{|k^2|},x,z \Big)} \nonumber \\
&\times& \Bigg\{ \delta (l_1^2) \delta (l_2^2)
-\delta (l_1^2+l_{1,\perp}^2) \delta (l_2^2)
-\delta (l_1^2) \delta (l_2^2+l_{2,\perp}^2)
+\delta (l_1^2+l_{1,\perp}^2) \delta (l_2^2+l_{2,\perp}^2) \Bigg\} \nonumber \\
&=& \frac{1}{4} \frac{\pi^{1-\epsilon}}{\Gamma (1-\epsilon)}
\frac{\pi^{\frac{1}{2}-\epsilon}}{\Gamma (\frac{1}{2}-\epsilon)}
\int_0^{Q^2} d|k^2| |k^2|^{1-2 \epsilon} x^{\epsilon} \int_0^1 dv v^{-\epsilon}
\int_0^1 dw \left( w \tilde{w} \right)^{-\epsilon} \int_0^{\pi} d\theta
\sin^{-2 \epsilon} \theta \nonumber \\
&\times& \Bigg[ \tilde{x}^{1-2 \epsilon} \tilde{v}^{-\epsilon} 
f\Big( \frac{\tilde{x}}{x} v w (1-v \tilde{x}),\frac{l_{1,\perp}^2}
{|k^2|v \tilde{x}},\frac{\tilde{x}\tilde{v}}{1-v \tilde{x}} 
\left( r_1^2 + r_2^2 -2 r_1 r_2 \cos \theta \right),
\frac{l_{2,\perp}^2}{2|k^2|\tilde{x}\tilde{v}},x,1-v\tilde{x} \Big) \nonumber\\
&&-\frac{\tilde{x}^{-\epsilon}}{\tilde{v}} f\Big( \frac{vw}{x},\frac{w}{x},
\tilde{x} \left( r_3^2 + r_4^2 -2 r_3 r_4 \cos \theta \right), 
\frac{l_{2,\perp}^2}{2|k^2|\tilde{x}},x,1 \Big) \nonumber \\ 
&&-\frac{\tilde{x}^{-\epsilon}}{\tilde{v}} f\Big( \frac{w \tilde{x}}{x},
\frac{w}{x},r_5^2 + r_6^2 -2 r_5 r_6 \cos \theta, \frac{\tilde{w}\tilde{v}}
{2x},x,x \Big) \nonumber \\
&&+ \frac{\delta (1-x)}{\tilde{v}} \int_0^1 d\rho \, \rho^{-2+\epsilon} 
\tilde{\rho}^{-\epsilon} f\Big( \frac{w \tilde{\rho}}{\rho}, 
\frac{w}{\rho},r_7^2 + r_8^2 -2 r_7 r_8 \cos \theta,
\frac{\tilde{w}\tilde{v}}{2},1,1 \Big) \Bigg] \; ,
\end{eqnarray}
where
\begin{eqnarray}
&& \tilde{x} = 1-x \;, \;\;\;  \tilde{v} = 1-v \;, \;\;\;  
\tilde{w} = 1-w \;, \;\;\;  \tilde{\rho} = 1-\rho \;, \nonumber \\
&&r_1=\sqrt{\tilde{w}} \;, \;\;\; r_2 = \sqrt{\frac{v w \tilde{v}\tilde{x}^2}
{x}} \; ,  \nonumber \\
&&r_3=\sqrt{\tilde{w}} \;, \;\;\; r_4 = \sqrt{\frac{v w \tilde{x}}{x}} 
\; ,  \nonumber \\
&&r_5=\sqrt{\frac{w\tilde{x}}{x}} \;, \;\;\; r_6 = \sqrt{v \tilde{w}} \; ,  
\nonumber \\
&&r_7=\sqrt{\frac{w\tilde{\rho}}{\rho}} \; , \;\;\; r_8= \sqrt{v \tilde{w}} 
\; .
\end{eqnarray}
The meaning of the arguments of the function $f$ in the various parts
of the phase space can be seen from the first line of~(\ref{b5}).
We note that the last part of the phase space in~(\ref{b5})
is proportional to $\delta (1-x)$; it corresponds to the contribution where
both gluons are axial ghosts.

We do not list the results for the numerous different phase space 
integrals one encounters. We just mention that one frequently needs
the integral   
\begin{eqnarray}
\int_0^{\pi} d\theta \frac{\sin^{-2 \epsilon}\theta}{r_a^2+r_b^2-2 r_a r_b
\cos \theta} = B\left( \frac{1}{2}-\epsilon,\frac{1}{2} \right) 
&\Bigg[& \frac{1}{r_a^2} \; {}_2 F_1 
\left( 1,1+\epsilon,1-\epsilon;\frac{r_b^2}{r_a^2} \right)
\Theta (r_a^2-r_b^2 ) \nonumber \\
&+& \frac{1}{r_b^2} \; {}_2 F_1 
\left( 1,1+\epsilon,1-\epsilon;\frac{r_a^2}{r_b^2} \right)
\Theta (r_b^2-r_a^2 ) \Bigg] \; , \nonumber \\
\end{eqnarray}
where ${}_2 F_1 (a,b,c;z)$ denotes the hypergeometric function~\cite{grad}. 
To expand in $\epsilon$, the relations 
\begin{eqnarray}
{}_2 F_1 (1,1+\epsilon,1-\epsilon;z) &=& (1-z)^{-1-2\epsilon} \; 
{}_2 F_1 (-\epsilon,-2\epsilon,1-\epsilon;z) \; , \\
{}_2 F_1 (a \epsilon,b \epsilon,1-c \epsilon;z)
&=&1+a b \epsilon^2 \Big( {\rm Li}_2 (z)+\epsilon c {\rm Li}_3 (z) +
\epsilon (a+b+c) {\rm S}_{1,2} (z) \Big) + {\cal O}(\epsilon^2) \nonumber 
\end{eqnarray}
are useful, where~\cite{dd}
\begin{equation}
{\rm Li}_3 (z) \equiv \int_0^1 \frac{\ln y \ln (1-z y)}{y} dy \; , \;\;\;
{\rm S}_{1,2} (z) \equiv \frac{1}{2} \int_0^1 \frac{\ln^2 (1-z y)}{y} dy \; .
\end{equation}
\newpage
\section*{Appendix C: Imaginary parts of loop integrals}
\appendix
\setcounter{equation}{0}
\renewcommand{\theequation}{C.\arabic{equation}}
To see how to extract the imaginary part of a loop integral, let us 
go back to our example in~(\ref{a4}) for the case $q=l$ there.
The integration over the Feynman parameter $s$ in~(\ref{a4}) is trivial and 
can be done immediately. Rather than performing straightaway the integration 
over $t$ to get the general result of the integral for arbitrary
$l^2$, it is more convenient to include the $\tau$ integration of~(\ref{ps11})
in the calculation and carry it out first:
\begin{eqnarray} \label{c1}
\int d\tau \tau^{-\epsilon} I (n,l)&=&
\frac{-i\Gamma(1+\epsilon)}{16\pi^2\epsilon} \frac{1}{[nl]} 
\Big( \frac{4 \pi}{|k^2|} \Big)^{\epsilon} (1-x)^{-\epsilon} x^{\epsilon}
\int_0^1 dt \, t^{-1-\epsilon} \int_{\tau_{min}}^{\tau_{max}} d\tau 
\tau^{-\epsilon} \nonumber \\ 
&\times&\Bigg[ ( \tau (1-t x)-1)^{-\epsilon} -
(1-t)^{-\epsilon} (\tau-1)^{-\epsilon} \Bigg] \; ,
\end{eqnarray}
where we have used the definitions in~(\ref{kdef}). To result in an imaginary
part\footnote{We obviously do not take into account the overall factor 
$i$ here.}, the limits for the $\tau$ integration have to be chosen 
in such a way that those terms in~(\ref{c1}), that are raised to the 
power $-\epsilon$, become negative, i.e., $0<\tau<1/(1-t x)$ for the 
first term in square brackets in~(\ref{c1}) and $0<\tau<1$ for the 
second. The $\tau$ integrations become trivial then and lead to 
simple beta--functions. Afterwards, the $t$ integration can be performed;
the result is given in Tab.~5 where we also list other integrals
that we encountered. As can be seen from Tab.~5, we also needed some
integrals with an extra factor $\tau$ or $(1-\tau)$ in the numerator.
We do not consider the covariant integrals in Tab.~5 since in their 
case the extraction of the imaginary part is rather straightforward.

As we discussed in Sec.~2.2 (see Eq.~(\ref{ip})), we have terms 
$\sim 1/l^2 \sim 1/(1-\tau)$ in the calculation, resulting from the 
propagator of the gluon running into the loop, and to be treated according 
to the principal value prescription. Therefore, we will also need integrals
like~(\ref{c1}) with an extra factor $1/(1-\tau)$ in the integrand. 
Such integrals are in general much more difficult to calculate. 
A typical integral needed is 
\begin{equation}
{\rm PV} \Bigg[ \int_0^{1/(1-t x)} \frac{d\tau}{1-\tau} \Bigg] 
= \ln \Bigg( \frac{1-t x}{tx} \Bigg) \; .
\end{equation}
The integrals with an extra denominator $1-\tau$ are also collected
in Tab.~5. 

We finally emphasize that Table~5 does {\em not} contain all terms $\sim
\epsilon$, but is in general correct only to ${\cal O} (1)$. The only
${\cal O}(\epsilon)$ terms that are fully accounted for in Tab.~5
are those $\sim \zeta (2)$. We have not consistently determined the other
contributions $\sim \epsilon$, because this is a very hard task. 
Therefore, since terms like $(1-x)^{-1-\epsilon}$ will lead to further 
pole factors $\sim \delta (1-x)/\epsilon$ via Eq.~(\ref{del}), we will not 
be able to calculate the {\em finite} amount of $\delta (1-x)$ in the final 
result for the $C_F N_C$ part of the two--loop splitting function, except for
the contributions $\sim \zeta (2) \delta (1-x)$. However, the expressions 
in Tab.~5 are sufficient for checking graph by graph the cancellation of 
all {\em pole} terms proportional to $\delta (1-x)$, i.e., for proving the 
finiteness of the 2PI kernels at $x=1$ in the ML prescription.
\newpage
\begin{table}[ht]
\begin{center}
\renewcommand{\arraystretch}{1.6}
\begin{tabular}{|c|c|c|} \hline \hline
$-{\cal IP} \Big[ \frac{1}{i\pi} \int d^d r/(2\pi)^d \Big]$ & 
$\int d\tau \tau^{-\epsilon} \tau^{\alpha} (1-\tau)^{\beta}$ & 
$\int d\tau \frac{\tau^{-\epsilon}}{1-\tau}$
\\ \hline \hline
$\frac{pn}{r^2 (l-r)^2 [nr]}$ & $\frac{1}{\tilde{x}} \Big[
( 1+2 \epsilon) \frac{1-\tilde{x}^{-\epsilon}}{\epsilon}
+ \epsilon \zeta (2) \Big] $ &  
$\frac{1}{2\epsilon^2} \tilde{x}^{-1-\epsilon} (1-\epsilon^2 \zeta (2))$ 
\\& $(\alpha =0,\beta=0)$ & $+\frac{1}{\tilde{x}} ({\rm Li}_2 (x)-
\zeta (2) )$ \\ \hline
$\frac{(r^2-2 nl \; n^* r) \; pn}{r^2 (l-r)^2 [nr]}$ & does not occur &
$ k^2 \Big[ -\frac{1}{\epsilon} -1-\frac{1}{x} +\ln x 
$ \\ & & $-\ln \tilde{x} (1-\frac{1}{x}) -
2 \tilde{x}^{-\epsilon} \Big]$  \\ \hline
$\frac{pn}{(p-r)^2 (k-r)^2 [nr]}$ & $-\tilde{x}^{-1-\epsilon} 
\ln x$ & $\tilde{x}^{-1-\epsilon} 
\Big[ \frac{\ln{x}}{\epsilon}+\frac{1}{2}\ln^2{x}$ \\ 
& $(\alpha=0,\beta=0)$ & 
$-2 {\rm Li}_2(\tilde x)+\epsilon \tilde{x} \zeta (2) \Big]$ \\ \hline 
$\frac{pn}{(p-r)^2 (l-r)^2 [nr]}$ & $-\frac{1}{\tilde{x}} 
\; \; \; \; \; (\alpha=0,\beta=0)$ & 
does not occur \\ \hline
$\frac{pn}{(k+r)^2 (l-r)^2 [nr]}$ & $\frac{1}{\tilde{x}} \Big[ 
\frac{1}{\epsilon} +2 +\ln x \Big]$ & 
$(1-\tilde{x}^{-\epsilon}) (\frac{1}{\epsilon^2} -\zeta (2))
-\frac{\ln x}{\epsilon}$ \\
& $(\alpha=0,\beta=0)$
& $+\zeta(2)-2 {\rm Li}_2 (\tilde{x})-\frac{1}{2} \ln^2 x$ \\ \hline 
$\frac{pn}{r^2 (p-r)^2 (k-r)^2 [nr]}$ & $\frac{1}{k^2} \tilde{x}^{-\epsilon}
\Big[ -\frac{1}{\epsilon}+\frac{2x}{\tilde x}\ln{x} +\epsilon \zeta (2) \Big]$
& $\frac{1}{k^2} \tilde{x}^{-\epsilon} \Big[\frac{1}{\epsilon^2}
+\frac{2 \ln x}{\epsilon}$ \\ 
& $(\alpha=0,\beta=0)$ & $ - 4 {\rm Li}_2 (1-x) -\zeta (2) \Big] $ \\ \hline 
$\frac{pn}{r^2 (k+r)^2 (l-r)^2 [nr]}$ & $\frac{1}{k^2 \tilde{x}}
\Big[ -(1-\tilde{x}^{-\epsilon}) (\frac{1}{\epsilon^2} -\zeta (2))$&
$\frac{1}{k^2 \tilde{x}}\Big[ \frac{3}{2} \tilde{x}^{-\epsilon} 
(\frac{1}{\epsilon^2} -\zeta (2))$ \\
& $-{\rm Li}_2 (x)-\ln x \ln \tilde{x} \Big]$ & 
$+2 ({\rm Li}_2 (x)-\zeta (2)) \Big] $\\
& $(\alpha=0,\beta=0)$ & \\
& $\frac{x}{k^2 \tilde{x}^2}\Big[ \frac{1}{\epsilon} +2 
-2 \ln \tilde{x} (1-\frac{1}{x})+\ln x\Big] $ & \\
& $(\alpha=1,\beta=0) $ & \\ \hline
$\frac{pn}{r^2 (p-r)^2 (l-r)^2 [nr]}$ & $\frac{x}{k^2 \tilde{x}}
\Big[ -(1-\tilde{x}^{-\epsilon}) (\frac{1}{\epsilon^2} -\zeta (2))+
\frac{\ln x}{\epsilon}$ & does not occur \\
& $+2 {\rm Li}_2 (x)+\frac{1}{2} \ln^2 x-3 \zeta (2) \Big]$ & \\ 
& $(\alpha =0,\beta=0)$ & \\
& $\frac{x}{k^2 \tilde{x}^2}\Big[ -\frac{\tilde{x}}{\epsilon} -2+3 x
+2 \tilde{x} \ln \tilde{x}$ &\\
& $- (1-3 x) \ln x \Big]\;\;\;\; (\alpha=0,\beta=1)$ & \\ \hline
\end{tabular}
\vspace*{0.3cm}
\caption{\sf Imaginary parts `${\cal IP}$' of loop integrals for the ML
prescription, after integration over the variable $\tau$ of the LO phase 
space~(\ref{ps11}). As before we have 
defined $\tilde{x}=1-x$. The integrals are in general only correct to
${\cal O}(1)$; see text. Note that terms $\sim (1-x)^{-\epsilon}$
must not be expanded in $\epsilon$, as further pole terms can arise via 
Eq.~(\ref{del}). We have dropped the ubiquitous factor 
$1/16\pi^2 \left( 4\pi/|k^2| \right)^{\epsilon} \Gamma(1-\epsilon)/
\Gamma(1-2 \epsilon)$.}
\vspace*{-0.5cm}
\end{center}
\end{table}
\newpage
\section*{Appendix D: Mellin moments}
\appendix
\setcounter{equation}{0}
\renewcommand{\theequation}{D.\arabic{equation}}
The Mellin--moments of a function $f(x)$ are defined by 
\begin{equation}
f^n \equiv \int_0^1 dx x^{n-1} \, f(x) \; .
\end{equation}
As a result, the moments of a convolution $f\otimes g$ (see Eq.~(\ref{conv1}))
become the product of the moments of $f$ and $g$:
\begin{equation}
(f\otimes g)^n = f^n \; g^n \; .
\end{equation}

The moments of~(\ref{conv}) are easily obtained using the formulae
in the appendix of~\cite{aem}. To invert the moments of the product 
$f^n g^n$ back to $x$--space, one needs some further moment expressions.  
Everything can be derived from the relations in~\cite{aem}, and from
\begin{eqnarray}
\int_0^1 dx \; 
x^{n-1} \ln^2 (1-x) &=& \frac{1}{n} \Big( S_1^2 (n)+S_2(n) \Big) 
\; , \nonumber \\ 
\int_0^1 dx \; x^{n-1} \Bigg[ \frac{\ln^2 (1-x)}{1-x} \Bigg]_+ 
&=& \frac{1}{n} \Big( S_1^2 (n)+S_2(n) \Big) 
-\frac{1}{3} S_1^3 (n) -S_1 (n) S_2 (n) -\frac{2}{3} S_3 (n) \; , \nonumber \\ 
\int_0^1 dx \; x^{n-1} {\rm Li}_2 (x) &=& \frac{1}{n} \zeta (2) - 
\frac{1}{n^2} S_1 (n) \; , \nonumber \\ 
\int_0^1 dx \; x^{n-1} 
\ln x \ln (1-x) &=& \frac{1}{n} \Big( S_2 (n) - \zeta (2) 
\Big) + \frac{1}{n^2} S_1 (n) \; , \\
\int_0^1 dx \; x^{n-1} \frac{\ln x \ln (1-x)}{1-x} &=& 
\Big( S_2 (n) -\zeta (2) \Big) \Big( \frac{1}{n} -S_1 (n) \Big) +
\frac{1}{n^2} S_1 (n) -S_3 (n) +\zeta (3) \; , \nonumber 
\end{eqnarray}
where
\begin{equation}
S_k (n) \equiv \sum_{j=1}^n \frac{1}{j^k} \; .
\end{equation}
\newpage
\section*{Appendix E: Two--loop integrals}
\appendix
\setcounter{equation}{0}
\renewcommand{\theequation}{E.\arabic{equation}}
For the calculation of the $C_F T_f$ part of the two--loop quark self--energy 
we need some integrals with an extra non--integer power of $-r^2$ in the integrand, 
where $r$ is the loop momentum. Making use of the identities
\begin{eqnarray}
\frac{1}{a^\alpha b} &=& \alpha \int_0^1 dx \frac{x^{\alpha -1}}
{\left[ a x+b (1-x) \right]^{\alpha +1}} \nonumber \; , \\
\frac{1}{a^\alpha bc} &=& \alpha (\alpha +1) \int_0^1 dx \int_0^{1-x} dy
\frac{x^{\alpha -1}}{\left[ a x+b y +c (1-x-y) \right]^{\alpha +2}} \; ,
\end{eqnarray}
one obtains rather easily:
\begin{eqnarray} \label{last1}
\int \frac{d^d r}{(2 \pi)^d} \frac{\left( -r^2 \right)^{-\epsilon}}
{(p-r)^2} &=& \frac{i}{16 \pi^2} (4 \pi)^{\epsilon} 
\left( -p^2 \right)^{1-2 \epsilon} \frac{\epsilon \Gamma (2 \epsilon)}
{\Gamma(1+ \epsilon)}\frac{\Gamma(1- \epsilon) \Gamma(1-2 \epsilon)}
{\Gamma(3-3 \epsilon)} \; , \\
\int \frac{d^d r}{(2 \pi)^d} \frac{\left( -r^2 \right)^{-\epsilon}}
{r^2 (p-r)^2} &=& \frac{i}{16 \pi^2} (4 \pi)^{\epsilon} 
\left( -p^2 \right)^{-2 \epsilon} \frac{\Gamma(2 \epsilon)}{\Gamma(1+ \epsilon)}
\frac{\Gamma(1- \epsilon) \Gamma(1-2 \epsilon)}{\Gamma(2-3 \epsilon)} \; , \\
\int \frac{d^d r}{(2 \pi)^d} \frac{\left( -r^2 \right)^{-\epsilon}}
{r^2 (p-r)^2 [nr]} &=& \label{last}
\left( -p^2 \right)^{-\epsilon} \left( 1+\epsilon^2 \zeta (2) \right) 
\int \frac{d^d r}{(2 \pi)^d} \frac{1}{r^2 (p-r)^2 [nr]} \; ,
\end{eqnarray}
where the integral on the right--hand--side of~(\ref{last}) was
determined in App.~A and is actually finite. For the ML prescription 
we therefore do not need the integral on the left--hand--side 
of~(\ref{last}); however, we will see below that the integral is 
divergent for the PV prescription. Also note that the integral 
in~(\ref{last1}) vanishes if the 
factor $\left( -r^2 \right)^{-\epsilon}$ is not present.

Finally, for the PV prescription one obtains for the integral in~(\ref{last}):
\begin{equation}
\int \frac{d^d r}{(2 \pi)^d} \frac{\left( -r^2 \right)^{-\epsilon}}
{r^2 (p-r)^2 [nr]} = \frac{i}{16 \pi^2} (4 \pi)^{\epsilon} 
\left( -p^2 \right)^{-2 \epsilon} \frac{1}{2\epsilon pn}
\left[ I_0+\epsilon \zeta (2) -2 \epsilon I_1 \right] +{\cal O} (\epsilon) \; ,
\end{equation}
while 
\begin{equation}
\int \frac{d^d r}{(2 \pi)^d} \frac{1}{r^2 (p-r)^2 [nr]} = 
\frac{i}{16 \pi^2} (4 \pi)^{\epsilon} 
\left( -p^2 \right)^{-\epsilon} \frac{1}{\epsilon pn}
\left[ I_0+\epsilon \zeta (2) -\epsilon I_1 \right] +{\cal O} (\epsilon) \; .
\end{equation}
Here
\begin{equation} \label{i1}
I_1 \equiv \int_0^1 \frac{u \ln u}{u^2+\delta^2} du \; .
\end{equation}
\newpage

\end{document}